\documentclass[reprint,APS,PRB,twocolumn,superscriptaddress]{revtex4}
\usepackage{url}
\usepackage{amsmath}
\usepackage{mathtools}
\usepackage{bbold}
\usepackage{mathrsfs}

\usepackage{graphicx}
\usepackage[caption=false]{subfig}
\usepackage{epstopdf} 
\usepackage{wrapfig}
\usepackage{braket}
\usepackage{xcolor}
\usepackage{comment}
\usepackage{lipsum}
\usepackage{cancel}
\usepackage{hyperref}
\usepackage{mathdots}

\usepackage{tikz}
\usepackage{circuitikz}
\usetikzlibrary{chains,fit,shapes}
\usetikzlibrary{cd} %
\usetikzlibrary{shapes.geometric}
\usetikzlibrary{decorations.markings}
\usetikzlibrary{decorations.pathmorphing,patterns}
\usetikzlibrary{decorations}
\usetikzlibrary{calc}
\usetikzlibrary{intersections}
\usetikzlibrary{arrows}
\usetikzlibrary{matrix}
\usetikzlibrary{positioning}
\usetikzlibrary{automata}

\newcommand{\RNum}[1]{\uppercase\expandafter{\romannumeral #1\relax}}

\newcommand{\mathU}{\mathbb{U}}
\newcommand{\mathH}{\mathbb{H}}

\newcommand{\CD}{\scriptscriptstyle \mathrm{CD}}
\newcommand{\opt}{\scriptscriptstyle \mathrm{opt}}
\newcommand{\LO}{\scriptscriptstyle \mathrm{LO}}
\newcommand{\btheta}{\boldsymbol{\theta}}

\def\ud{\mathrm{d}}

\newcommand{\nep}{\mathrm{e}}

\newcommand{\defuguale}{\stackrel{\mathrm{def}}{=}}

\newcommand{\x}{{\mathbf x}}

\newcommand{\A}{{\mathbf A}}

\newcommand{\C}{{\mathbf C}}
\newcommand{\rmC}{{\mathrm C}}
\newcommand{\G}{{\mathbf G}}

\newcommand{\target}{\mathrm{\scriptscriptstyle targ}}
\newcommand{\drive}{\mathrm{\scriptscriptstyle drive}}

\newcommand{\Heis}{\mathrm{\scriptscriptstyle H}}

\newcommand{\transpose}{\mathsf{T}}

\newcommand{\Nbasis}{\mathrm{N_c}}

\newcommand{\U}{{\mathbf U}}
\newcommand{\V}{{\mathbf V}}

\newcommand{\calA}{{\mathcal{A}}}

\newcommand{\Tr}{{\rm Tr}}

\newcommand{\taucr}{\tau^*}

\newcommand{\prm}{\mathrm{p}}

\newcommand{\opc}[1]{{\hat{c}^{\phantom \dagger}}_{#1}}
\newcommand{\opcdag}[1]{{\hat{c}^{\dagger}}_{#1}}

\newcommand{\Gammaopup}{\hat{\Gamma}^{+}}
\newcommand{\Gammaopdown}{\hat{\Gamma}^{-}}
\newcommand{\Gammaopbf}{\hat{\mathbf{\Gamma}}}
\newcommand{\PauliSigma}{\hat{\sigma}}

\newcommand{\Ho}{\hat{H}}
\newcommand{\Uo}{\hat{U}}

\newcommand{\mathGamma}{\mathbb{\Gamma}}
\newcommand{\opbfPsidag}[1]{{\widehat{\mathbf{\Psi}}^{\dagger}}_{#1}}

\newcommand{\Nsites}{\mathrm{N}}
\newcommand{\Nfermions}{\mathrm{\hat{N}_F}}

\newcommand{\fin}{\mathrm{fin}}

\newcommand{\Ptrot}{\mathrm{P}}

\newcommand{\opbfPsi}[1]{{\widehat{\mathbf{\Psi}}^{\phantom \dagger}}_{#1}}
\newcommand{\Hc}{\mathrm{H.c.}}

\newcommand{\deltat}{\Delta_t}

\newcommand{\gs}{\mathrm{\scriptstyle gs}}
\newcommand{\ex}{\mathrm{\scriptstyle ex}}

\graphicspath{{./}{./Fig/}}

\begin{document}
\author{Vincenzo Roberto Arezzo}
\email[Email (Corresponding Author): ]{varezzo@sissa.it}
\affiliation{SISSA, Via Bonomea 265, I-34136 Trieste, Italy}

\author{Kiran Thengil}
\affiliation{SISSA, Via Bonomea 265, I-34136 Trieste, Italy}

\author{Giuseppe E. Santoro}
\affiliation{SISSA, Via Bonomea 265, I-34136 Trieste, Italy}
\affiliation{International Centre for Theoretical Physics (ICTP), P.O.Box 586, I-34014 Trieste, Italy}

\title{Continuous-time quantum control across an exponentially small bottleneck in a frustrated Ising ring model}

\begin{abstract}
Continuous-time Quantum Annealing (QA) is a strategy for preparing the ground state of nontrivial many-body systems. In its standard form, the dynamics is generated by a time-dependent interpolation between a simple driving Hamiltonian and the target problem Hamiltonian, usually implemented through a linear schedule. This approach faces the crucial bottleneck of small spectral gaps, which may require exponentially long annealing times to ensure adiabaticity.  Here, we show how to implement quantum control over the annealing schedule in a frustrated Ising ring, one of the simplest models exhibiting an exponentially small bottleneck gap. By optimizing smooth continuous-time annealing schedules with a dressed-CRAB approach, and using a digitized representation of the dynamics to efficiently evaluate gradients, we construct protocols that strongly outperform standard fixed schedules. The optimized dynamics bypasses the bottleneck through a strongly nonadiabatic mechanism, leading to efficient ground-state preparation despite the exponentially small minimum gap. In particular, the annealing time required to reach a fixed residual-energy threshold is found to grow linearly with system size rather than exponentially. We further examine a lowest-order variational counter-diabatic correction and find that, once schedule optimization is allowed, it does not lead to any improvement. 
\end{abstract}

\maketitle

\section{Introduction}

Quantum annealing (QA) is a computational paradigm in which the solution of an optimization problem is encoded in the ground state of a target Hamiltonian $\Ho_{\target}$, while the system is initialized in the easily prepared ground state $|\psi_0\rangle$ of a driving Hamiltonian $\Ho_{\drive}$~\cite{finnila_quantum_1994,kadowaki1998quantum,Santoro_SCI02,Santoro_2006,Albash_RMP18}. 
These two terms enter in the total Hamiltonian through a time-dependent interpolation
\begin{equation}
    \Ho(t)=s(t)\Ho_{\target}+\big(1-s(t)\big)\Ho_{\drive} \,,
\end{equation}
and the state of the system $|\psi(t)\rangle$ obeys the time-dependent Schr\"odinger equation $i\partial_t|\psi(t)\rangle=\Ho(t)|\psi(t)\rangle$ starting from $|\psi(t=0)\rangle=|\psi_0\rangle$.
Here the function $s(t)$ --- henceforth, the {\em schedule} --- satisfies $s(0)=0$ and $s(\tau)=1$, where $\tau$ is the total evolution (or annealing) time.
In the adiabatic limit, for sufficiently large annealing time $\tau$, the evolved state remains close to the instantaneous ground state of $\Ho(t)$ and reaches the desired solution at the end of the protocol. 
The performance of standard QA is therefore strongly affected by the spectral properties of the interpolating Hamiltonian $\Ho(t)$, and in particular by its minimum instantaneous energy gap encountered during the evolution.
As demonstrated long ago by Roland \& Cerf~\cite{Roland_PRA2002}, a QA implementation of the Grover search problem~\cite{Grover_PRL1997} would fail to provide any quantum speed-up unless the schedule $s(t)$ is suitably improved over the standard linear choice $s(t)=t/\tau$. 
A recent detailed discussion of various possible strategies for improvement of the schedule, notably when passing through a second-order quantum critical point, is given in Ref.~\cite{Gabrarits_PRA2025}.
The situation becomes dramatically more severe when the spectral gap of the instantaneous Hamiltonian becomes exponentially small with the system size, as the annealing time required to maintain adiabaticity would become exponentially large 
~\cite{Caneva_PRB2007,Knysh_NatComm16,Knysh_PRA2020,Zamponi_QA:review}.

A natural way to overcome this limitation is to abandon the requirement of strict adiabaticity and regard the annealing schedule $s(t)$ itself as a control field. 
Rather than following a fixed schedule, such as a simple linear ramp $s(t)=t/\tau$, one can optimize $s(t)$ to minimize the final average target energy $\langle \psi(\tau)| \Ho_{\target}|\psi(\tau)\rangle$ at a given finite time $\tau$.
This point of view connects QA with the broader framework of quantum optimal control, where suitably designed time-dependent protocol fields can exploit nonadiabatic transitions as a resource rather than treating them only as errors to be suppressed~\cite{Matsuura_PRA2021,Cote_2023,Quiroz_PRA2019,Lucignano_PRA2022,Passarelli_PRB2019,Caneva_PRA2011,Montangero_dCRAB_PRA2015}. 
A complementary strategy is provided by counter-diabatic driving, where additional driving terms 
are added to the Hamiltonian to suppress transitions out of the instantaneous ground state~\cite{Berry_2009,KOLODRUBETZ20171,OdelinShortcuts2019,TorronteguiShortcuts2013}. 
Related ideas have also been incorporated into variational and digitized quantum optimization schemes, inspired by the Quantum Approximate Optimization Algorithm (QAOA)~\cite{Farhi_arXiv2014,HegadeDigitized2022,ChandaranaDigitized2022,IevaCounterdiabatic2023,Lubasch_PRXQ2024}.

In many-body systems, however, both schedule optimization and counter-diabatic driving are challenging. 
The control field is a continuous function of time, and the Hilbert-space dimension grows exponentially with the number of degrees of freedom. 
It is therefore important to identify nontrivial models in which the interplay between exponentially small gaps, nonadiabatic dynamics, and optimal control can be studied in detail.

In this work we focus on a frustrated Ising ring, introduced in Ref.~\cite{Knysh_PRA2020} and further studied in Refs.~\cite{Cote_2023,wang2025exponential,GrabaritsFighting2025}. 
The model consists of an Ising chain with an odd number of sites, two weak ferromagnetic bonds, and a single antiferromagnetic bond inducing frustration. 
Despite its apparent simplicity, this system displays a spin-glass bottleneck characterized by an exponentially small avoided crossing close to the end of the annealing path. 
As a consequence, standard adiabatic QA requires exponentially long times to reliably prepare the (simple) ferromagnetic ground state of the model.  
At the same time, after a Jordan-Wigner mapping into spinless fermions~\cite{jordan_uber_1928,mbeng2024quantum}, the model is quadratic and can be simulated efficiently. 
This makes it an ideal test bed for investigating whether optimized continuous-time schedules can bypass the exponentially small bottleneck.

Previous work on this model has shown that the bottleneck can be mitigated by exploiting diabatic dynamics~\cite{Cote_2023}. 
More recently, optimal control and QAOA-like approaches have shown that in digitized dynamics the scaling can be substantially improved by using nonadiabatic protocols and optimized variational parameters~\cite{wang2025exponential,Arezzodigital2025,werner_arxiv2026}. 
Counter-diabatic strategies have also been proposed as a way to fight exponentially small gaps in this class of problems~\cite{GrabaritsFighting2025,Kiran_in_preparation}. 
Here we take a complementary continuous-time perspective and ask whether smooth schedule optimization alone can reproduce, and further clarify, the nonadiabatic mechanisms responsible for this improvement. 

The main contribution of this work is the construction and analysis of smooth continuous-time schedules $s(t)$ showing that the exponentially small bottleneck can be bypassed by a controlled diabatic pathway.
We optimize smooth annealing schedules $s(t)$ using a Chopped RAndom Basis (CRAB) strategy~\cite{Montangero_dCRAB_PRA2015,Caneva_PRA2011,Doria_PRL2011,koch_quantum_2022}, in which the schedule $s(t)$ is expanded over a finite set of Fourier components with randomized frequencies. 

Our results show that schedule optimization alone strongly outperforms standard fixed schedules, including the traditional linear ramp, but also cubic and sinusoidal ramps with vanishing time derivatives at the beginning and at the end of the annealing, inspired by the adiabatic theorem~\cite{Morita_JMP2008}.  
The improvement is associated with a strongly nonadiabatic mechanism: during the evolution, population weight is transferred away from the instantaneous ground state and is later brought back to the target ground state near the end of the protocol. 
In this sense, the optimized protocol does not overcome the exponentially small bottleneck by slowing down the dynamics, but by exploiting controlled diabatic transitions across it.

We also investigate whether this optimized strategy can be further improved by adding a lowest-order variational counter-diabatic correction. 
Counter-diabatic driving is designed to suppress transitions out of the instantaneous ground state by adding a term proportional to the adiabatic gauge potential~\cite{Berry_2009,KOLODRUBETZ20171}. 
For the frustrated ring, the lowest-order local approximation to this gauge potential can be written explicitly and remains quadratic in the fermionic representation. 
However, when combined with schedule optimization, we find that this correction does not produce any improvement over the optimized schedule alone. 

The remaining paper is organized as follows. 
Section~\ref{sec:model} introduces the frustrated-ring model and its Jordan--Wigner fermionic representation.  
Section~\ref{sec:time_evolution} describes the time-discretized evolution, the connection with a QAOA-like {\em Ansatz}, and the CRAB optimization of the continuous schedule. 
In Sec.~\ref{sec:results} we present the optimized residual energies and compare them with various fixed schedules. 
Section~\ref{sec:populations} analyzes the populations of the instantaneous eigenstates and identifies the nonadiabatic mechanism underlying the optimized protocols. 
Section~\ref{sec:counter-diabatic_improvements} discusses the addition of the lowest-order counter-diabatic term and compares its performance with schedule optimization alone. 
Section~\ref{sec:conclusions} summarizes our conclusions and discusses possible future directions.
Finally, a few Appendices discuss some technical aspects of our study. 

\section{Model} \label{sec:model} 
We start by presenting the model we investigated. 
We consider the following transverse-field quantum Ising chain   
\begin{equation} \label{eqn:ann_hamiltonian_s}
\begin{aligned} 
\Ho(t) &= s(t) \Ho_{\target} + (1-s(t)) \Ho_{\drive}  \vspace{3mm} \\
\Ho_{\target} &\equiv \Ho_z = -\sum_{j=1}^{\Nsites} J_j \PauliSigma^z_j\PauliSigma^z_{j+1}
 \vspace{3mm}\\
\Ho_{\drive} &\equiv \Ho_x = -h \sum_{j=1}^{\Nsites} \PauliSigma^x_j
\end{aligned}
\end{equation}
where $s(t)$ denotes a generic real interpolation schedule between the transverse-field term $\Ho_x$, with a constant transverse field $-h$, attained when $s=0$, and a longitudinal nearest-neighbor coupling $\Ho_z$, attained when $s=1$.  
This model was proposed in Ref.~\cite{Knysh_PRA2020} and studied in Refs.~\cite{Cote_2023,wang2025exponential,GrabaritsFighting2025}.
The couplings $J_j$ between adjacent spins are:
\begin{equation}
  J_{j} =
    \begin{cases}
      J_w & \text{if $j=(\Nsites\pm 1)/2$}\\
      -J_f & \text{if $j=\Nsites$}\\
      J & \text{otherwise}
    \end{cases}       \;.
\end{equation}
Here, $J$ is the ferromagnetic coupling between most sites, $J_w$ denotes the weaker ferromagnetic coupling between the two central sites at $j=(\Nsites\pm 1)/2$, and $J_f$ is the single frustrating (antiferromagnetic) coupling at site $\Nsites$. 
(In the notation of Refs.~\cite{Knysh_PRA2020} and \cite{Cote_2023}, $J_f=J_R$ and $J_w=J_L$.)
We choose $0<J_f<J_w<J$, specifically requiring that $JJ_f>J_w^2$, which leads to the spin-glass bottleneck regime studied in Ref.~\cite{Knysh_PRA2020}. 
For the rest of this work we take, following Ref.~\cite{Cote_2023}, $J=1$, our unit of energy, $J_w=0.5$ and $J_f=0.45$ or $J_f=0.2501$. 
The ground state of $\Ho_z$ is the ferromagnetic state with energy $E_{\gs}
=-(\Nsites-3)J-2J_w+J_f$; the first excited state has two domain walls, at a central site 
$j=(\Nsites+1)/2$ and at $j=\Nsites$, and is separated from the ground state by an energy gap $\Delta_1\equiv E_{\ex}-E_{\gs}=2(J_w-J_f)$.

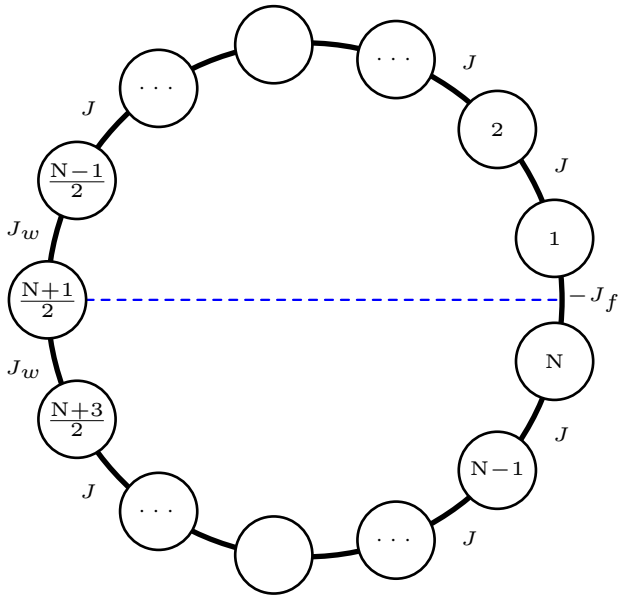
\begin{figure}[!htp]
\centering
\begin{tikzpicture}[scale=1.7,line cap=round,line width=2pt]
\filldraw [fill=black!0!white] (0,0) circle (2cm);
\draw [line width = 0.3mm, draw=blue, dashed] (-2.2,0) -- (2.0,0) node[right, black] {};
\foreach \x [evaluate=\x as \angle using (\x-0.5)*360/13] in {1,...,13}
{
\draw[line width=1pt,fill=white] ({2*cos(\angle)},{2*sin(\angle)}) circle (3mm);
}
\node[draw=none,font=\tiny,text=black,scale=1.5] at ({2*cos(0.5*360/13)},{2*sin(0.5*360/13)}) {$1$};
\node[draw=none,font=\tiny,text=black,scale=1.5] at ({2*cos(1.5*360/13)},{2*sin(1.5*360/13)}) {$2$};
\node[draw=none,font=\tiny,text=black,scale=1.5] at ({2*cos(2.5*360/13)},{2*sin(2.5*360/13)}) {$\cdots$};
\node[draw=none,font=\tiny,text=black,scale=1.5] at ({2*cos(4.5*360/13)},{2*sin(4.5*360/13)}) {$\cdots$};
\node[draw=none,font=\tiny,text=black,scale=1.5] at ({2*cos(5.5*360/13)},{2*sin(5.5*360/13)}) {$\scriptstyle{\frac{\Nsites-1}{2}}$};
\node[draw=none,font=\tiny,text=black,scale=1.5] at ({2*cos(6.5*360/13)},{2*sin(6.5*360/13)}) {$\scriptstyle{\frac{\Nsites+1}{2}}$};
\node[draw=none,font=\tiny,text=black,scale=1.5] at ({2*cos(7.5*360/13)},{2*sin(7.5*360/13)}) {$\scriptstyle{\frac{\Nsites+3}{2}}$};
\node[draw=none,font=\tiny,text=black,scale=1.5] at ({2*cos(8.5*360/13)},{2*sin(8.5*360/13)}) {$\cdots$};
\node[draw=none,font=\tiny,text=black,scale=1.5] at ({2*cos(-0.5*360/13)},{2*sin(-0.5*360/13)}) {$\Nsites$};
\node[draw=none,font=\tiny,text=black,scale=1.5] at ({2*cos(-1.5*360/13)},{2*sin(-1.5*360/13)}) {$\scriptstyle{\Nsites-1}$};
\node[draw=none,font=\tiny,text=black,scale=1.5] at ({2*cos(-2.5*360/13)},{2*sin(-2.5*360/13)}) {$\cdots$};
\node[draw=none,font=\tiny,text=black,scale=1.5] at ({2.25*cos(5*360/13)},{2.25*sin(5*360/13)}) {$J$};
\node[draw=none,font=\tiny,text=black,scale=1.5] at ({2.25*cos(6*360/13)},{2.25*sin(6*360/13)}) {$J_w$};
\node[draw=none,font=\tiny,text=black,scale=1.5] at ({2.25*cos(7*360/13)},{2.25*sin(7*360/13)}) {$J_w$};
\node[draw=none,font=\tiny,text=black,scale=1.5] at ({2.25*cos(8*360/13)},{2.25*sin(8*360/13)}) {$J$};
\node[draw=none,font=\tiny,text=black,scale=1.5] at ({2.25*cos(2*360/13)},{2.25*sin(2*360/13)}) {$J$};
\node[draw=none,font=\tiny,text=black,scale=1.5] at ({2.25*cos(1*360/13)},{2.25*sin(1*360/13)}) {$J$};
\node[draw=none,font=\tiny,text=black,scale=1.5] at ({2.25*cos(0)},{2.25*sin(0)}) {$\scriptstyle{-J_f}$};
\node[draw=none,font=\tiny,text=black,scale=1.5] at ({2.25*cos(-1*360/13)},{2.25*sin(-1*360/13)}) {$J$};
\node[draw=none,font=\tiny,text=black,scale=1.5] at ({2.25*cos(-2*360/13)},{2.25*sin(-2*360/13)}) {$J$};
\end{tikzpicture}
\caption{The Ising frustrated ring model with an odd number of sites $\Nsites$. The model has a reflection symmetry around the axis indicated by the dashed blue line.}
\label{fig:ring}
\end{figure}

We now use the Jordan-Wigner transformation \cite{mbeng2024quantum} to transform $\Ho(t)$ into a quadratic fermionic Hamiltonian: 
\begin{equation}
\begin{aligned}
\Ho_{\target} &\equiv \Ho_z = -\sum_{j=1}^{\Nsites} J_j 
\big( \opcdag{j} \opc{j+1} \,+\, \opcdag{j} \opcdag{j+1}  \,+\, {\Hc} \big) \vspace{3mm}\\
\Ho_{\drive} &\equiv \Ho_x = h \sum_{j=1}^{\Nsites} \big( \opcdag{j} \opc{j} - \opc{j} \opcdag{j} \big) 
\end{aligned}  \;,
\end{equation}
with boundary condition $\opc{\Nsites+1}=(-1)^{\prm+1} \opc{1}$,
where $\hat{c}^\dagger_j$ and $\hat{c}_j$ are fermionic creation and annihilation operators. 
Here, $\prm=0$ or $1$ denotes even or odd fermion parity sectors of the Hilbert space, $(-1)^{\prm}=\nep^{i\pi \Nfermions}$, where 
\[ \Nfermions=\sum_{j=1}^{\Nsites} \opcdag{j}\opc{j} \] 
is the total number of fermions.
The ground state of $\Ho_{x}$ for $h<0$, which we choose, is the fully occupied fermionic state 
\[ 
|\psi_0\rangle=\prod_{j=1}^{\Nsites} \opcdag{j}|0\rangle \;.
\]
This implies that the relevant fermion parity is $\prm=1$ when $\Nsites$ is odd. 
Hence, we fix $\Nsites$ to be odd and restrict our dynamics to the $\prm = 1$ fermion parity sector, where the fermions have periodic boundary conditions, $\hat{c}_{\Nsites+1} = \hat{c}_1$.

The couplings $J_j$ have a reflection symmetry $J_j=J_{\Nsites-j}$, see Fig.~\ref{fig:ring}, which can be exploited to simplify the computational task, as suggested in Ref.~\cite{GrabaritsFighting2025}. 
We summarize the basic mapping to exploit this reflection symmetry in Appendix \ref{app:reflection}.

In the strong frustration case $J_f\lesssim J_w$, this model shows two closing gaps between the ground state and the first excited state, as shown already in Ref.~\cite{wang2025exponential} (Fig.~2); the first one occurs at $s_c\approx 0.5$ and decays polynomially, as $1/\Nsites$: it is the critical gap separating the quantum paramagnetic phase ($s<s_c$) from the frustrated ferromagnetic phase ($s>s_c$); 
the second, at $s_b\approx 0.9$ (for $J_f=0.45$ and $J_w=0.5)$, is the exponentially small gap providing the spin-glass bottleneck discussed in Ref.~\cite{Knysh_PRA2020}.

\section{Time evolution and quantum control} 
\label{sec:time_evolution}
In this section we describe the methodology that we used to simulate continuous-time evolution. In continuous-time QA we have a time-dependent interpolating Hamiltonian:
\begin{equation}
\Ho(t) = s(t) \, \Ho_z + (1-s(t)) \, \Ho_x \;,
\label{eqn:ann_hamiltonian}
\end{equation}
where $s(0)=0$, $s(\tau)=1$, $\tau$ being the total annealing time. The system state $|\psi(t)\rangle$ will follow the Schr\"odinger equation
\begin{equation} \label{eqn:SE}
    i\hbar \frac{\ud }{\ud t} |\psi(t)\rangle = \Ho(t) |\psi(t)\rangle \;,
\end{equation}
with initial condition $|\psi(t=0)\rangle=|\psi_0\rangle$, where $|\psi_0\rangle$ is the ground state of $\Ho_x$.

Since the final quantum state $| \psi(t=\tau)\rangle$, and hence the average energy 
\begin{equation}
E^{\fin}(\tau)=\langle 
\psi(\tau)| \Ho_z | \psi(\tau)\rangle \;,
\end{equation} 
depend on $s(t)$ --- a continuous function of $t$ --- this optimal control in continuous time is a {\em functional optimization problem}. 
To bypass this difficulty, we first go towards a digitized QA dynamics~\cite{Nature_dQA,Mbeng_dQA_PRB2019,mbeng_quantum_2019,pecci2024beyond}. 
The idea is to discretize the dynamics into a large number $\Ptrot$ of very small time steps 
$\deltat = \tau/\Ptrot$. Setting $\hbar = 1$, we obtain
\begin{equation}
|\psi_{\Ptrot}\rangle  = \nep^{-i \deltat \Ho(t_{\Ptrot}) } \cdots 
\nep^{-i\deltat \Ho(t_{p})} \cdots
\nep^{-i\deltat \Ho(t_{1})} |\psi_0\rangle \;,
\end{equation}
where we choose the times $t_p$ in the middle of the corresponding intervals:
\begin{equation}
t_p = \big(p-\frac{1}{2}\big) \deltat  \hspace{5mm} p=1,\cdots, \Ptrot \;. 
\end{equation}
After this time-discretization step, it is possible to Trotterize the dynamics using the lowest-order Trotter splitting:
\begin{equation}
\nep^{-i\deltat (s_p\Ho_z+(1-s_p)\Ho_x)} \xrightarrow{\mathrm{Trotter}} 
\nep^{-i\theta^x_p \Ho_x} \nep^{-i\theta^z_p \Ho_z} \;,
\end{equation}
where $s_p=s(t_p)$ and
\begin{equation}
\theta^z_p = s_p \deltat \,, \hspace{10mm} 
\theta^x_p = (1-s_p) \deltat \;.
\end{equation}
For our simulations we choose $\deltat=\tau/\Ptrot=0.1$, namely $\Ptrot=10 * \mathrm{int}(\tau)$. This choice is justified in App.~\ref{app:trotter}, where the induced Trotter error is carefully examined.   This leads to the usual {\em Quantum Approximate Optimization Algorithm} (QAOA)~\cite{Farhi_arXiv2014} {\em Ansatz}:
\begin{equation}
|\psi_{\Ptrot}(\btheta)\rangle = 
\Uo_{\Ptrot}(\theta^x_\Ptrot,\theta^z_\Ptrot) \, 
\cdots \, \Uo_1(\theta^x_1,\theta^z_1) |\psi_0\rangle \;,
\end{equation}
where, for $p=1\cdots \Ptrot$:
\begin{equation}
\Uo_p(\theta^x_p,\theta^z_p) = 
\nep^{-i\theta^x_p \Ho_x} \nep^{-i\theta^z_p \Ho_z} \;.
\end{equation}
Next, we make an {\em Ansatz} for the schedule $s(t)$ 
in the spirit of the Chopped RAndom Basis (CRAB) approach of Ref.~\cite{Caneva_PRA2011}, based on the Fourier basis functions in the time-interval $[0,\tau]$. 
This can be done iteratively. 
Let $s^{(0)}(t)=t/\tau$ be the linear schedule. For $k\ge 1$, 
if $s^{(k-1)}(t)$ is the current schedule, then at iteration $k$ we define $s^{(k)}(t)$ as follows:
\begin{equation} \label{eqn:schedule_CRAB}
\begin{aligned}
s^{(k)}(t) = s^{(k-1)}(t) + \sin\left({\textstyle \frac{\pi}{\tau}}t\right) 
\Bigg( &\sum_{n=1}^{\Nbasis} {\text{C}^s_n \sin(\omega_n t)} \; + \\
& \sum_{n=1}^{\Nbasis} \text{C}^c_n 
\cos(\omega_n t) 
\Bigg) \;,
\end{aligned}
\end{equation}
i.e., we add a finite number, $2\Nbasis$, of Fourier components
with random frequencies $\omega_n$ to the previous iteration schedule.  
This scheme is known as {\em dressed}-CRAB~\cite{Montangero_dCRAB_PRA2015}, and is able to progressively improve the schedule by keeping the optimization problem finite-dimensional: at each step the goal is to optimize the $2\Nbasis$ variational parameters $\C=(\C^s,\C^c)$, with $\C^{s/c}=(\rmC^{s/c}_1,\cdots,\rmC^{s/c}_{\Nbasis})$, by minimizing the average final energy
\begin{equation} \label{eqn:E_dQA}
E_{\Ptrot}(\C) = \langle \psi_{\Ptrot}(\btheta(\C)) | \Ho_z | \psi_{\Ptrot}(\btheta(\C)) \rangle \;.
\end{equation}
The appropriate values of $\Nbasis$, and how to choose the corresponding frequencies $\omega_n$ was a matter that we addressed by numerical experiments. 
To fix the ideas, one could take $\Nbasis = 20$ for each dressed-CRAB iteration and select the dimensionless frequencies $x=\omega_n\tau/\pi$ from a Gamma probability distribution:
\begin{align}
\Gamma (x\ge 0; \alpha, \beta) = \frac{x^{\alpha-1} \nep^{-x/\beta}}{\beta^\alpha\Gamma(\alpha)} \;,
\end{align}
where $\alpha>0$ determines the scale at small values of $x$, and $\beta>0$ the overall exponential scale at large $x$. 
$\Gamma(\alpha)$ is the Euler Gamma function. 
This distribution is peaked at $x_{\max}=(\alpha-1)\beta$, and we used $\alpha=3/2$ as shape parameter. To reduce the dependence on a particular random frequency drawn, we repeat the frequency extraction $N_{\text{real}} = 20$ times and retain the set that gives the lowest optimized value of $E_{\Ptrot}(\btheta)$.
We typically did $k=2$ iterations of dressed-CRAB. 
In the first iteration, $k=1$, we choose $\beta=4$, so that the distribution is peaked at $x_{\max}=2$: this provides the first optimized schedule $s^{(1)}(t)$ with the most important (small frequency) Fourier components. 
Next, we performed a second iteration, $k=2$, with $\beta=8$, to give the final schedule $s^{(2)}(t)$ higher frequency components that improve the quality of the solution. 
Unless otherwise stated, all optimized schedules shown in the following are obtained using two dressed-CRAB iterations with \(\Nbasis=20\) Fourier components per iteration.  

The time-discretization and Trotterization of the dynamics is very useful in the solution of the finite-dimensional optimization problem, since it gives easy access to the analytical gradients of the function $E_{\Ptrot}(\btheta(\C))$, see Ref.~\cite{wang2025exponential} (Appendixes C and D), which were passed to the Broyden-Fletcher-Goldfarb-Shanno (BFGS) algorithm \cite{nocedal1999numerical} used to optimize the average energy $E_{\Ptrot}$. 

Alternatively, the gradients can be evaluated directly in continuous time,
for instance by augmenting the equations of motion with those for the
parameter derivatives, as in Ref.~\cite{Machnes2018}, or by differentiating
through the numerical ODE solver using automatic-differentiation techniques,
as in Ref.~\cite{Sauvage2022PRL}. Related approaches for analog quantum
dynamics are discussed in Appendix~C.3 of Ref.~\cite{Sauvage2025PRR}.

Once the optimal variational parameters $\C^{\opt}$ are obtained, in principle a full schedule $s^{\opt}(t)$ in continuous time is available, see Eq.~\eqref{eqn:schedule_CRAB}, and one might integrate the Schr\"odinger equation~\eqref{eqn:SE} using a fourth-order adaptive Runge-Kutta algorithm to calculate the final energy:
\begin{equation}
\label{eqn:rungekuttaenergy}
    E^{\fin}(t=\tau) = \langle \psi(\tau) | \Ho_z | \psi(\tau) \rangle \;.
\end{equation}
The comparison between $E_{\Ptrot}(\C^{\opt})$ and $E^{\fin}(\tau)$ is discussed in App.~\ref{app:trotter} and shows that the discrepancy between the two quantities decreases approximately as $1/\Ptrot^{2}$, confirming that the optimized protocols represent genuine continuous-time schedules for large enough $\Ptrot$.

\begin{figure*}[htp]
\centering
\includegraphics[width=160mm]{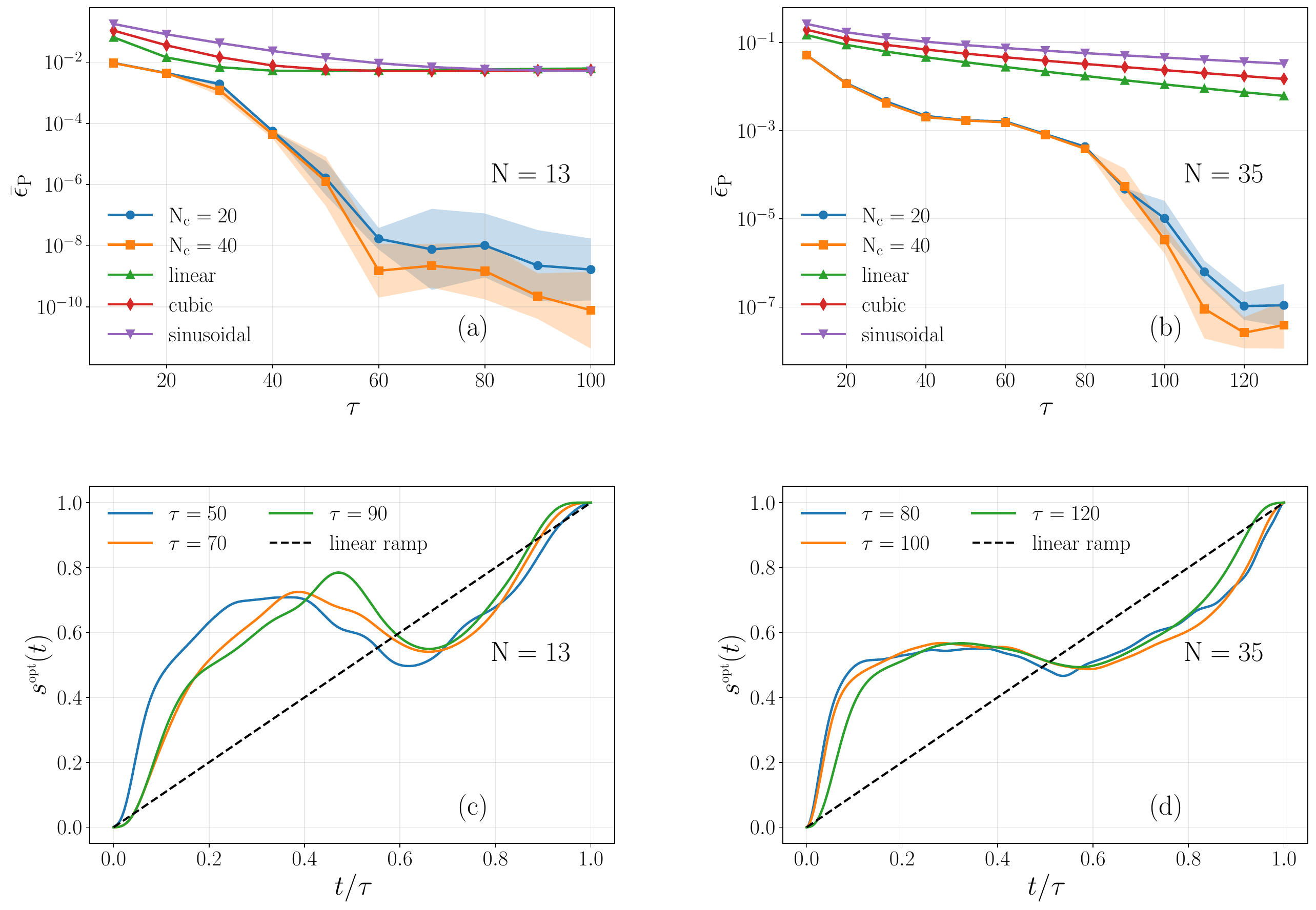}
\caption{Comparison between optimized continuous-time residual energies and fixed-schedule results for $\Nsites = 13$, $J_f = 0.45$ (a,c), and $\Nsites = 35$, $J_f = 0.45$ (b,d). Panels (a) and (b) show the geometric mean of the optimized residual energies $\bar{\epsilon}_{\Ptrot}(\tau)$ over $N_{\text{run}} = 10$ independent dressed-CRAB runs. The shaded regions indicate one logarithmic standard deviation. Panels (c) and (d) show representative optimized schedules $s^{\opt}(t)$ for the case $\Nbasis = 20$, and several values of $\tau$. 
}
\label{fig:residualenergyscalingN13}
\end{figure*}

\begin{figure}[htp]
\centering

\includegraphics[width=0.99\columnwidth]{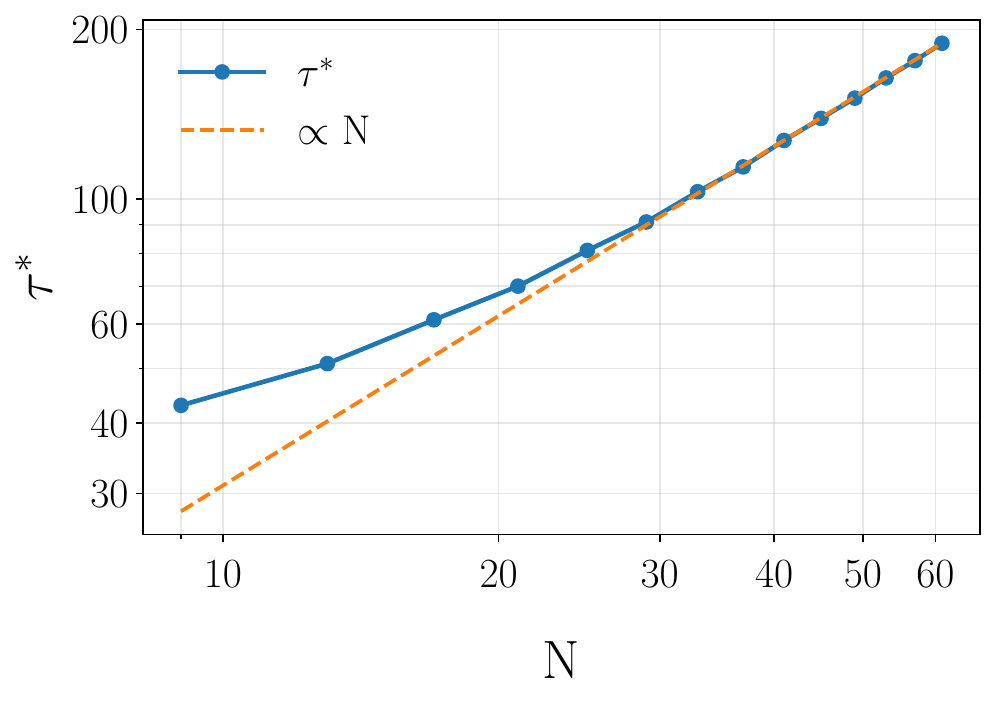}

\caption{
Threshold annealing time $\taucr$ extracted from the condition 
$\bar{\epsilon}_{\Ptrot}(\tau)<\epsilon^{\scriptscriptstyle\mathrm{threshold}}=10^{-6}$, as a function of the system size $\Nsites$. 
For each $\Nsites$, we scan the annealing time $\tau$ in steps of $\Delta\tau=1$ and define $\taucr$ as the smallest scanned value of $\tau$ for which the geometric mean of the optimized residual energy satisfies
$\bar{\epsilon}_{\Ptrot}(\tau)<\epsilon^{\scriptscriptstyle\mathrm{threshold}}$.
The data are shown in log-log scale. 
The dashed line is a linear-scaling guide to the eye, $\taucr\propto \Nsites$, normalized using the largest system size. 
The comparison suggests that, in the range of sizes explored here, the threshold time $\taucr$ is compatible with a linear scaling with $\Nsites$. All data are obtained with the default two-iteration dressed-CRAB setup described in Sec.~\ref{sec:time_evolution}.
}
\label{fig:tau_threshold_geom}
\end{figure}

\section{Results} \label{sec:results}
Here we present the results obtained by applying a dressed-CRAB control to the schedule $s(t)$ for the frustrated ring model.
The results are first compared to fixed schedule results, including the standard linear schedule $s^{\scriptscriptstyle\mathrm{lin}}(t)=t/\tau$, but also schedules with a vanishing derivative at $t=0$ and $t=\tau$, which should in principle achieve, based on the quantum adiabatic theorem~\cite{Morita_JMP2008}, a better asymptotic scaling as a function of $1/\tau$. Among these, we mention the cubic schedule~\cite{GrabaritsFighting2025}
\begin{equation}
s^{\scriptscriptstyle\mathrm{cub}}(t) = 
(t/\tau)^2\, \big( 3-2(t/\tau) \big) \;,
\label{cubicrampeqn}
\end{equation}
and the sinusoidal schedule~\cite{HegadeDigitized2022,IevaCounterdiabatic2023}
\begin{equation}
\label{sinrampeqn}
s^{\scriptscriptstyle\mathrm{sin}}(t) 
= \sin^{2}\left[\frac{\pi}{2}\sin^{2}\left(\frac{\pi t}{2 \tau}\right)\right] 
\;.
\end{equation}

To assess the quality of the solutions obtained, we will use, as a figure of merit, a rescaled residual energy, both in discrete-time and continuous-time dynamics:
\begin{equation} \label{eqn:res_energy}
\begin{aligned}
\epsilon_{\Ptrot}(\tau) 
&= \frac{E_{\Ptrot}(\C^{\opt}) - E_{\gs}}{\Nsites} \;, \\
\epsilon_{\text{ct}}(\tau) &= \frac{E^{\text{fin}}(\tau) - E_{\gs}}{\Nsites}
\end{aligned} \;,
\end{equation}
where $E_{\Ptrot}(\C^{\opt})$ is the optimal value of the expression in Eq.~\eqref{eqn:E_dQA}, and  $E^{\text{fin}}(\tau)$ the corresponding optimal continuous-time expression in Eq.~\eqref{eqn:rungekuttaenergy}.
Here $E_{\gs}=-(\Nsites-3)J-2J_w+J_f$ is the ground-state energy of the frustrated ring model with $\Nsites$ sites. 
In the following, we analyze only the discrete-time residual energy $\epsilon_{\Ptrot}(\tau)$, discussing the comparison with $\epsilon_{\text{ct}}(\tau)$ in App.~\ref{app:trotter}.

For each value of $\tau$ we performed $N_{\mathrm{run}}=10$ independent dressed-CRAB optimizations, corresponding to different random choices of the CRAB frequencies, and leading to different values of 
$\epsilon^{(r)}_{\Ptrot}(\tau)$, with $r=1\cdots N_{\mathrm{run}}$.
Since the residual energies are non-negative and span several orders of magnitude, we analyze their fluctuations on a logarithmic scale, defining their logarithmic mean and standard deviation as:
\begin{equation} \nonumber
\begin{aligned}
\label{eqn:geommeananddev}
\mu_{\log}(\tau) &= \frac{1}{N_{\mathrm{run}}}
    \sum_{r=1}^{N_{\mathrm{run}}} \log \epsilon^{(r)}_{\Ptrot}(\tau), \\
\sigma_{\log}(\tau) &=
    \sqrt{
    \frac{1}{N_{\mathrm{run}}-1}
    \sum_{r=1}^{N_{\mathrm{run}}}
    \left[
    \log \epsilon^{(r)}_{\Ptrot}(\tau)-\mu_{\log}(\tau)
    \right]^2
    } \;.
\end{aligned}
\end{equation}
The representative residual energy is then taken to be the geometric mean
\begin{equation}
\label{eqn:geommean}
    \bar{\epsilon}_{\Ptrot}(\tau)
    = \nep^{\mu_{\log}(\tau)} \,,
\end{equation}
while the error bars are obtained from a logarithmic standard deviation interval~\footnote{This is a natural choice for positive quantities spanning several orders of magnitude. It should not be understood as assuming that the residual energies $\epsilon^{(r)}_{\Ptrot}(\tau)$ are exactly log-normally distributed.},
\begin{equation}
\label{eqn:geomstddeviation}
\nep^{\mu_{\log}(\tau)-\sigma_{\log}(\tau)} \le \epsilon_{\Ptrot}(\tau) \le \nep^{\mu_{\log}(\tau)+\sigma_{\log}(\tau)} \;.
\end{equation}

\subsection{Optimized continuous-time schedule} \label{sec:controllability}
We start by presenting in Fig.~\ref{fig:residualenergyscalingN13} the optimized residual energies as a function of $\tau$, together with the corresponding optimal schedules $s^{\opt}(t)$, for two chain sizes $\Nsites=13$ and $\Nsites=35$, and for two different values of the CRAB-basis truncation, $\Nbasis=20$ and $40$. 
The residual-energy points correspond to the geometric mean $\bar{\epsilon}_{\Ptrot}(\tau)$, see Eq.~\eqref{eqn:geommean}, over $N_{\text{run}} = 10$ independent dressed-CRAB optimizations, while the shaded regions indicate one logarithmic standard deviation, as defined in Eq.~\eqref{eqn:geomstddeviation}.  
As expected, the residual energy decreases as $\tau$ increases. 
Figure \ref{fig:residualenergyscalingN13} also shows results obtained for three fixed schedules, namely linear $s^{\mathrm{lin}}=t/\tau$, cubic (see Eq.~\eqref{cubicrampeqn}), and sinusoidal (see Eq.~\eqref{sinrampeqn}). These fixed schedule results show that the dynamics remains trapped in a bottleneck-induced plateau over the range of $\tau$ explored:
Reaching vanishing residual energies with such fixed protocols would require much longer, indeed exponentially large, annealing times~\cite{GrabaritsFighting2025,wang2025exponential,Cote_2023}. 
By contrast, the optimized dressed-CRAB schedules produce a much faster reduction of the residual energy, indicating that a substantial improvement can already be achieved at moderate annealing times.

The comparison with fixed schedules also highlights an important point. Cubic and sinusoidal ramps have vanishing time derivatives at the boundaries of the protocol and are therefore expected, on the basis of the adiabatic theorem, to reduce boundary-induced diabatic errors with respect to a linear schedule~\cite{Morita_JMP2008}. 
This advantage, however, is only asymptotic, for $\tau\to \infty$. 
For the finite annealing times considered here, the residual energy is dominated by the bottleneck dynamics rather than by boundary effects. 
As a result, the smoother boundary behavior of the cubic and sinusoidal schedules does not provide any advantage, and these ramps perform even worse than the linear schedule.

To characterize more quantitatively the scaling of the control performance with system size, we extract for each $\Nsites$ a threshold annealing time $\taucr$, defined as the first value of $\tau$ --- technically related to a value of $\Ptrot=10  *  \mathrm{int}(\tau)$ --- for which the geometric mean of the residual energy over the different dressed-CRAB runs satisfies 
$\bar{\epsilon}_{\Ptrot}(\tau)<\epsilon^{\scriptscriptstyle\mathrm{threshold}}=10^{-6}$. 
The resulting $\taucr$ is shown in Fig.~\ref{fig:tau_threshold_geom}. 
In the range of sizes explored here, $\taucr$ grows linearly with $\Nsites$. 
This is an important indication that the optimized continuous-time control strategy avoids the exponentially large times associated with purely adiabatic evolution across the exponentially small spectral gap bottleneck.

\begin{figure*}[htp]
\centering
\includegraphics[width=0.88\textwidth]{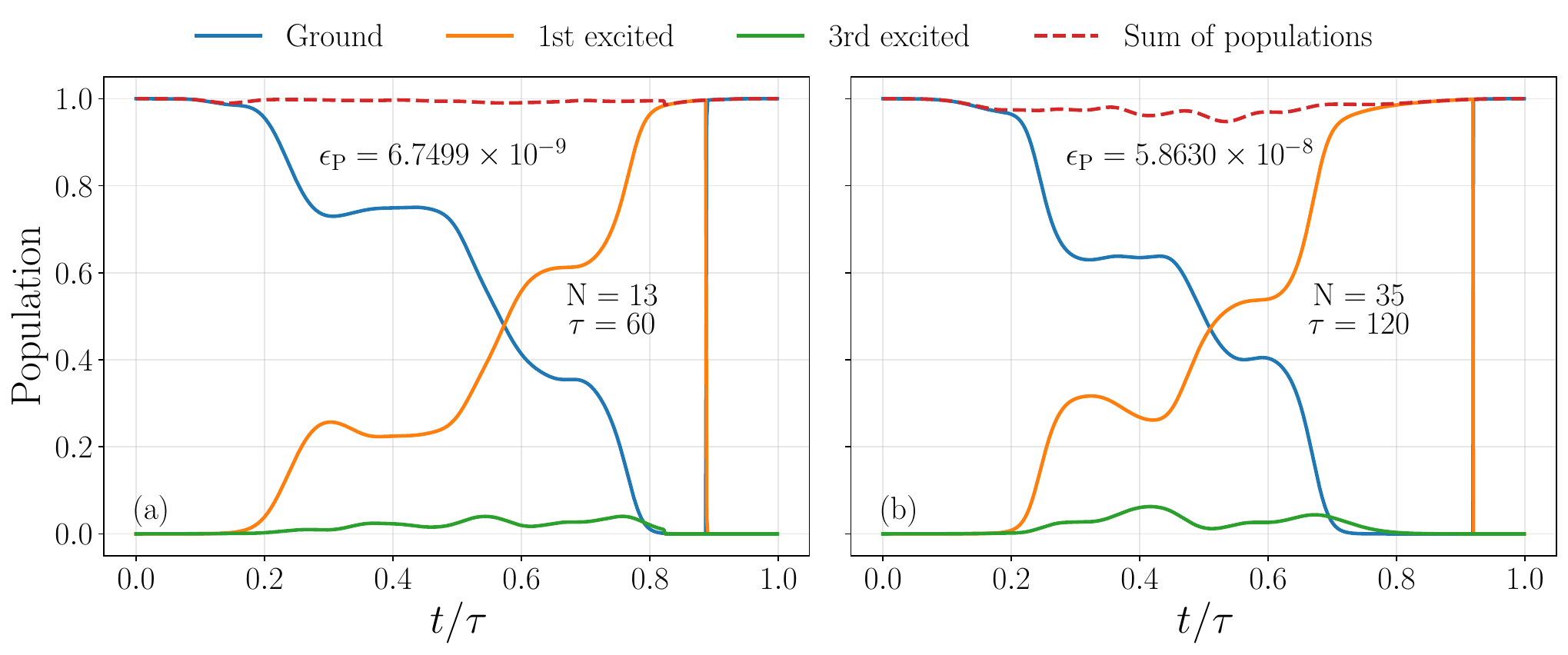}
\caption{
Populations of the low-lying instantaneous eigenstates during the optimized dynamics for two system sizes:
$\Nsites=13$ at $\tau=60$ (panel (a)) and $\Nsites=35$ at $\tau=120$ (panel (b)).
All data are obtained with the default two-iteration dressed-CRAB setup described in Sec.~\ref{sec:time_evolution}.
Notice the abrupt final population transfer for values of $t/\tau$ close to the bottleneck value of $s_b\approx 0.9$.
The final ground-state infidelities are
$1-p_0(\tau) \simeq 3.8\times 10^{-8}$ for $\Nsites=13$ and
$1-p_0(\tau) \simeq 9.6\times 10^{-7}$ for $\Nsites=35$.
Since both the initial state and $\Ho(s)$ are even under the reflection symmetry of Fig.~\ref{fig:ring}, only even eigenstates can be populated; the second excited state is odd and is therefore not shown.
}
\label{fig_notes:populations}
\end{figure*}

\subsection{Populations of the instantaneous eigenstates}
\label{sec:populations}
In order to understand the physical mechanism behind the optimized schedules, we analyze the time-dependent populations of the instantaneous eigenstates of the annealing Hamiltonian
\begin{equation}
    \Ho(s)=s\Ho_z+(1-s)\Ho_x \;.
\end{equation}
At each time $t$, we diagonalize $\Ho(s(t))$ and denote its instantaneous eigenstates by
$|\phi_n(s(t))\rangle$, ordered by increasing energy. 
The corresponding populations are defined as
\begin{equation}
    p_n(t)=\left|\langle \phi_n(s(t))|\psi(t)\rangle\right|^2 \,,
\end{equation}
where $|\psi(t)\rangle$ is the state evolved under the optimized continuous-time schedule.

For a perfectly adiabatic evolution, the population would remain concentrated in the instantaneous ground state throughout the whole protocol, namely $p_0(t)\simeq 1$ for all $t\in[0,\tau]$. 
The optimized protocols found by dressed-CRAB display a markedly different behavior. 
As shown in Fig.~\ref{fig_notes:populations}, the dynamics is strongly nonadiabatic: the system leaves the instantaneous ground state during the early stage of the evolution, developing a sizable population in the first excited state, and returns almost completely to the target ground state at the end of the evolution. This transfer is the mechanism through which the optimized schedule bypasses the exponentially small bottleneck gap: 
Once the first excited state is populated, the optimized protocol exploits the final avoided crossing associated with the exponentially small gap to transfer the population back to the instantaneous ground state. 
This allows the system to reach the target state.

The population analysis also clarifies why the optimized schedules outperform standard fixed schedules. 
Linear, cubic, or sinusoidal ramps are constrained to follow a predetermined path in time and therefore cannot efficiently coordinate the sequence of diabatic transitions required to return to the target ground state. 
An interesting physical insight into the mechanism of population transfer, through Landau-Zener-St\"uckelberg interference, is discussed in Ref.~\cite{werner_arxiv2026}.

\subsection{Variational counter-diabatic driving}
\label{sec:counter-diabatic_improvements}
The basic idea of counter-diabatic driving~\cite{KOLODRUBETZ20171}, also known as transitionless quantum driving~\cite{Berry_2009}, consists in adding extra terms to the driving Hamiltonian that exactly cancel the nonadiabatic transitions inevitably occurring when the annealing time $\tau$ is finite: 
\begin{align} \label{eqn:cd_driving}
i \frac{\ud}{\ud t} |\psi(t)\rangle = 
\underbrace{\Big( s \Ho_z + (1-s) \Ho_x + \dot{s} \hat{\calA}(s) \Big)}_{\Ho_{\CD}(s,\dot{s})}
|\psi(t)\rangle \;,
\end{align}
where we set $\hbar = 1$. These extra terms involve $\dot{s}$ multiplying the so-called {\em adiabatic gauge potential} $\hat{\calA}(s)$, a highly non-local operator for which variational optimization schemes have been introduced in Ref.~\cite{KOLODRUBETZ20171}. 
To lowest order (LO), an analytical expression for the optimal variational adiabatic gauge potential is given by~\cite{KOLODRUBETZ20171}:
\begin{align}
\hat{\calA}(s) \stackrel{\mathrm{lowest\, order}}{\approx} \hat{\calA}^{\LO}(s) 
&= -i \alpha(s) \big[ \Ho(s),\partial_s \Ho(s) \big] \nonumber \\
&= -i \alpha(s) \big[ \Ho_x,\Ho_z \big] \;,
\end{align}
or, upon translating into Jordan-Wigner fermions
~\cite{mbeng2024quantum}:
\begin{equation}
\begin{aligned}
\hat{\calA}^{\LO}(s) &=  -i \alpha(s)[\Ho_x , \Ho_z] \\
&=  4i \alpha(s) \sum_{j=1}^{\Nsites} J_j 
\Big( \opcdag{j} \opcdag{j+1} - \opc{j+1} \opc{j} \Big) \;.
\end{aligned}
\end{equation}
Here the optimal function $\alpha(s)$ involves traces of Pauli
operators, and can be calculated explicitly: 
\begin{align} \label{eqn:alpha_s}
\alpha(s) &= \frac{  \Tr \big(i[\Ho_x,\Ho_z]\big)^2}{\Tr\big( (1-s) [\Ho_x,[\Ho_x,\Ho_z]] + s [\Ho_z,[\Ho_x,\Ho_z]] \big)^2} \nonumber \\
&=\frac{\sum_j J_j^2}{16 (1-s)^2 \sum_j J_j^2 + 4s^2( 3 \sum_j J_j^2 J_{j+1}^2 + \sum_j J_j^4 ) } \;.
\end{align}
Details are given in App.~\ref{app:counter-diabatic}.

The schedule $s(t)$ appearing in Eq.~\eqref{eqn:cd_driving} is still left unspecified, and in principle one might optimize it. The idea is that, for the {\em exact} $\Ho_{\CD}(s,\dot{s})$, any schedule would do the job, as the system remains by construction in the instantaneous ground state manifold. 
When the lowest-order-local adiabatic gauge potential $\hat{\calA}^{\LO}(s)$ is adopted, the driving Hamiltonian is
\begin{equation}
\Ho_{\CD}^{\LO}(s,\dot{s}) = s \Ho_z + (1-s) \Ho_x + \dot{s} \hat{\calA}^{\LO}(s) \;,
\end{equation}
however, excitations out of the ground state are possible, and a judicious (optimal) choice of the schedule $s(t)$ should exist.

\begin{figure*}[htp]
\centering
\includegraphics[width=160mm]{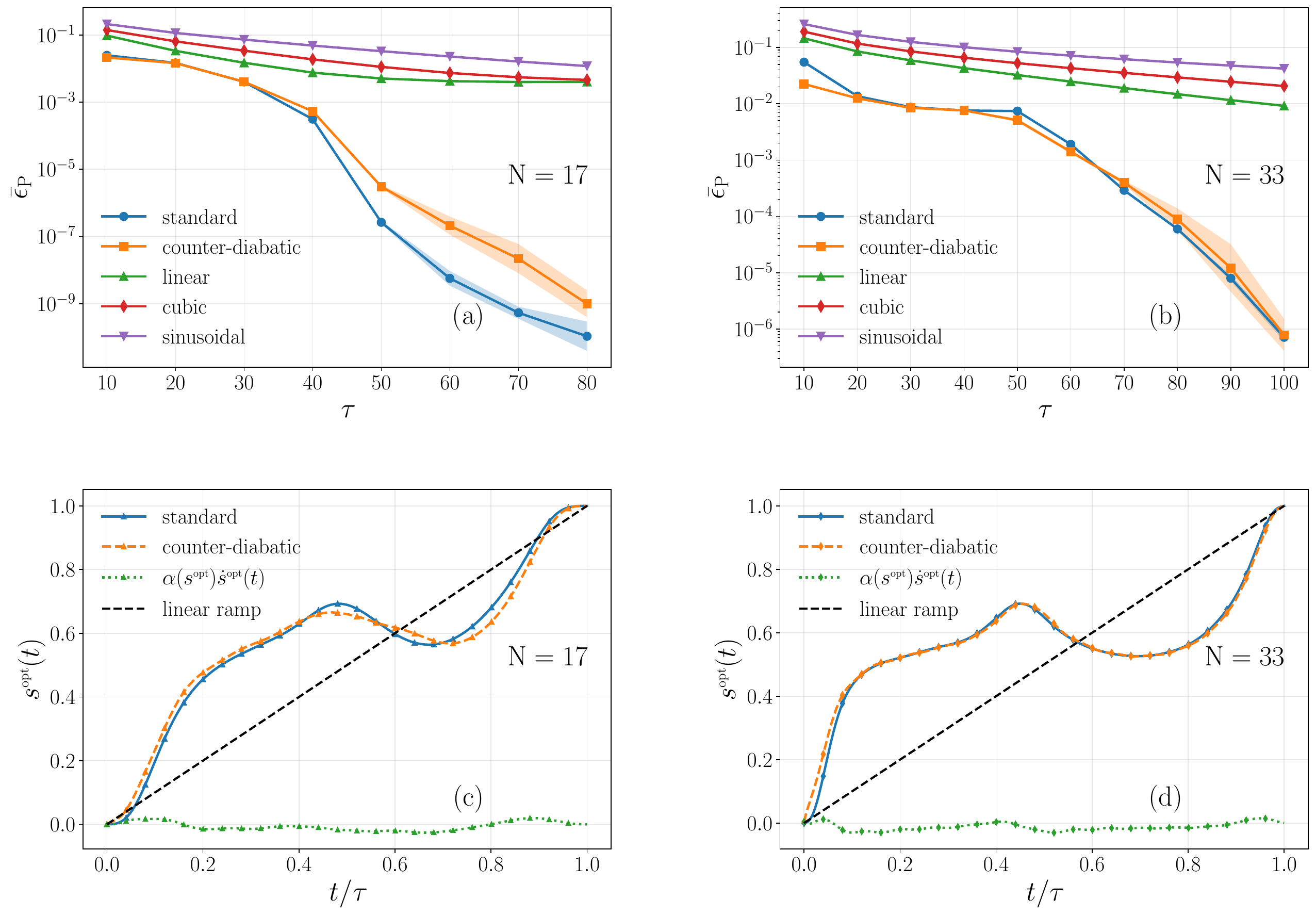}
\caption{
Comparison of the optimized residual energies obtained with the standard and counter-diabatic protocols, together with the results for fixed linear, cubic, and sinusoidal schedules.
The data refer to $J_f=0.2501$ and $\Nsites=17$ in panels (a,c), and to $J_f=0.2501$ and $\Nsites=33$ in panels (b,d).
The residual-energy points show the geometric mean $\bar{\epsilon}_{\Ptrot}(\tau)$ over $N_{\text{run}} = 10$ independent dressed-CRAB runs, while the shaded regions indicate one logarithmic standard deviation.
The bottom panels compare the optimized schedules for $\tau=80$ ($\Nsites=17$) and $\tau=100$ ($\Nsites=33$) and show the corresponding profile of $\alpha(s)\dot{s}(t)$ entering the counter-diabatic correction.
Both the standard and counter-diabatic schedules are optimized using the same default dressed-CRAB setup described in Sec.~\ref{sec:time_evolution}.
}
\label{fig:counter_N17}
\end{figure*}

To compare our results with those of Ref.~\cite{GrabaritsFighting2025} we show in Fig.~\ref{fig:counter_N17} residual energies for the cases of $\Nsites = 17$ and $\Nsites = 33$, with a small frustration bond $J_f = 0.2501$.  Our results show that adding the lowest-order variational counter-diabatic term does not improve the final residual energy. 
This behavior can be understood in physical terms. The optimized schedules found through dressed-CRAB do not follow an adiabatic strategy: population is first transferred to the first excited state in the first part of the evolution and then returns to the target ground state near the end of the evolution by exploiting the small avoided crossing at $s_b \approx 0.9$ (see sec.~\ref{sec:populations}). 
By contrast, the role of the counter-diabatic term is to suppress transitions and keep the system as close as possible to the instantaneous ground state of the annealing Hamiltonian.  
Therefore, in this frustrated Ising ring model, it is not surprising that combining optimal control with a counter-diabatic term does not yield any advantage.

In Ref.~\cite{GrabaritsFighting2025}, a cubic schedule augmented by a lowest-order counter-diabatic term was found to slightly improve the annealing dynamics of the frustrated ring, mainly at shorter driving times. Our results for $\Nsites = 17$ and $\Nsites = 33$, however, indicate that significant improvements can be obtained only through control optimization.

\section{Conclusions and perspectives} 
\label{sec:conclusions}
In this work we investigated continuous-time quantum control in a frustrated Ising ring model displaying an exponentially small spectral gap. 
This model provides a setting in which standard adiabatic quantum annealing is expected to require exponentially long times, while the quadratic fermionic structure allows for efficient numerical simulations at relatively large system sizes.

We optimized smooth annealing schedules using a dressed-CRAB approach. 
The continuous-time control problem was represented through a finely digitized dynamics, allowing us to use analytical gradients and efficient BFGS optimization.
This strategy allowed us to compare optimized protocols with standard fixed schedules, such as linear, cubic, and sinusoidal ramps.

Our main result is that continuous-time schedule optimization can reduce the final residual energy by several orders of magnitude with respect to fixed schedules.
Cubic and sinusoidal schedules have vanishing derivatives at the boundaries and are therefore expected to suppress boundary-induced diabatic corrections~\cite{Morita_JMP2008}. Instead, we found that they perform worse than the linear schedule in the time regime considered here, leading to the conclusion that the necessary ingredient to bypass the exponentially small gap is the protocol optimization. 
 
The optimized schedules generate a strongly nonadiabatic dynamics: the state leaves the instantaneous ground state, develops significant population in the first excited instantaneous eigenstate, and is finally transferred back to the target ground state near the end of the evolution using the avoided crossing at 
$s_b \approx 0.9$. 
We found that the annealing time required to reach a fixed residual-energy threshold grows approximately linearly with the system size over the range investigated, rather than following the exponential scaling expected from a purely adiabatic passage through the bottleneck gap. This scaling is not the same as the one found in~\cite{wang2025exponential,Arezzodigital2025} for the minimum number of QAOA steps $\Ptrot^{\text{cr}}$ necessary to reach controllability in the digital setting: $\Ptrot^{\text{cr}} \propto \Nsites^2$. 
This finding, however, is not in contrast with our result, since we are considering a continuous-time setting with a finite fixed accuracy threshold $\epsilon^{\scriptscriptstyle\mathrm{threshold}}=10^{-6}$, different from the one imposed for full controllability (nominally, $\epsilon^{\scriptscriptstyle\mathrm{threshold}}=0$).
Our linear scaling also compares favorably with the $\propto \Nsites^2$ scaling of the median annealing time found in Ref.~\cite{Cote_2023} for optimized piecewise-linear schedules ($\Nsites\le 39$), and with the $\propto \Nsites^{\alpha}$ scaling, $\alpha\approx 1.2$--$1.4$, reported in Ref.~\cite{werner_arxiv2026} for schedules exploiting Landau-Zener-St\"uckelberg interference.

We also studied the effect of adding a lowest-order variational counter-diabatic correction. 
Although such a term is designed to suppress  transitions and can be written explicitly for the present free-fermionic model, we found that it does not yield any improvement once the schedule is already optimized. 
This result is consistent with the population analysis in Sec.~\ref{sec:populations}: the optimized schedule relies on a diabatic mechanism where the instantaneous ground-state population is appropriately initially depleted, while the counter-diabatic terms try to do exactly the opposite. 

Several directions of investigation remain open.
A first natural extension concerns the structure of the optimized schedules below the controllability threshold. In Ref.~\cite{Brady_PRL2021}, it was shown that, in related quantum annealing and QAOA control problems, the optimal finite-time protocols may not be purely smooth, but can develop bang-bang features localized in the beginning and the end of the evolution. In a detailed analysis of the uniform Ising chain, to be presented in Ref.~\cite{Arezzo_inpreparation}, we show that this type of boundary bang-bang behavior systematically emerges below the controllability threshold. It would therefore be interesting to extend the present dressed-CRAB ansatz by allowing for boundary bang segments, and to test whether such hybrid protocols can further improve the performance in the frustrated-ring model. 

A second perspective concerns the optimization strategy and the relationship between digitized and continuous-time control. The Trotterized representation used here provides an efficient way to evaluate gradients,
but alternative approaches based on direct continuous-time integration~\cite{Machnes2018,Sauvage2022PRL,Sauvage2025PRR}, could avoid the Trotter splitting error. 
Conversely, higher-order Baker--Campbell--Hausdorff~\cite{rossmann2006lie} corrections to the digitized representation would generate effective commutator terms, such as $[\hat{H}_x,\hat{H}_z]$. Including such corrections could provide a more accurate bridge between the digitized optimization procedure and the target continuous-time dynamics~\cite{Wurtz_2022}.

Finally, a further important question is whether similar diabatic mechanisms can be explicitly demonstrated in more complex many-body models with small spectral gaps, equally hard for QA. In particular, it remains unclear whether the coherent back-and-forth population transfer observed here can survive in genuinely interacting systems, where the spreading of excitations over a larger portion of the many-body Hilbert space may hinder their coherent transfer back to the ground state.
Extending the analysis to non-integrable spin chains, higher-dimensional frustrated models, or models that are characterized by more than one avoided crossing, would shed more light into the generality of the mechanisms analyzed in the present paper: an interesting route, in that respect, is to study models that can be tackled with Matrix Product States \cite{Lami_SciPost2023} and other Tensor Network techniques~\cite{Collura2025}, where quantum control techniques have not been attempted so far, to the best of our knowledge.

\begin{acknowledgments}
G.E.S. and K.T. acknowledge financial support from PNRR MUR Project No.~PE0000023-NQSTI, funded by the European Union -- NextGenerationEU. 
G.E.S. acknowledges financial support from PRIN 2022H77XB7 of the Italian Ministry of University and Research, funded by the European Union -- NextGenerationEU, and from the QuantERA II Programme STAQS project that has received funding from the European Union's H2020 research and innovation programme under Grant Agreement No.~101017733.
\end{acknowledgments}

\section*{Data availability}
The data that support the findings of this article, together with a Jupyter notebook that reproduces the figures, are openly available~\cite{Arezzo_Zenodo2026}.

\begin{widetext}
\appendix
\section{The reflection symmetry mapping} \label{app:reflection}

To exploit reflection symmetry of the couplings $J_j$, we perform the transformation introduced in Ref.~\cite{GrabaritsFighting2025}. 
We define a new set of $2\Nsites$ Dirac fermions $\hat{\Gamma}^{\pm}_{j=1\cdots \Nsites}$, appropriately defined on odd and even sites as follows: 
\begin{align} \label{eq:gammaoddj}
\hat{\Gamma}^{\pm}_{2j-1} &= 
\frac{i}{2} \Big(\pm (\opc{j} + \opcdag{j}) +(\opc{\Nsites+1-j} - \opcdag{\Nsites+1-j}) \Big) 
\end{align}
for $j=1,\cdots, (\Nsites+1)/2$, and 
\begin{align} \label{eq:gammaevenj}
\hat{\Gamma}^{\pm}_{2j} &= 
\frac{i}{2}
\Big((\opc{j} -\opcdag{j}) 
\pm (\opc{\Nsites +1 -j} + \opcdag{\Nsites+1-j}) \Big) \;,
\end{align}
for $j=1,\cdots, (\Nsites-1)/2$. 
Observe that 
\begin{equation}
\Gammaopup_j = \left(\Gammaopdown_j\right)^{\dagger} \;,
\end{equation} 
and that they obey standard fermionic anticommutation relations~\cite{GrabaritsFighting2025}:
\begin{equation}
\begin{aligned}
\{\Gammaopdown_j,\Gammaopup_{j'}\} = \delta_{j,j'} \;, \hspace{5mm}
\{\Gammaopup_j,\Gammaopup_{j'}\} = 
\{\Gammaopdown_j,\Gammaopdown_{j'}\} = 0 \;.
\end{aligned}
\end{equation}
The expressions for  $\hat{\Gamma}^{\pm}_{j}$ can be cast into a convenient Nambu matrix form which encodes a unitary transformation of the original Nambu fermions into new symmetry-adapted Nambu operators:
\begin{equation} \label{eqn:Gamma_vs_Psi}
\Gammaopbf
\equiv \left(
\begin{array}{c}
\Gammaopdown_{1} \\
\Gammaopdown_{2} \\
\vdots \\ 
\Gammaopdown_{\Nsites} \\
\Gammaopup_{1} \\
\Gammaopup_{2} \\
\vdots \\
\Gammaopup_{\Nsites}
\end{array}
\right) 
= \mathU_{\Gamma}
\left(
\begin{array}{c}
\opc{1} \\
\opc{2} \\
\vdots \\
\opc{\Nsites} \\
\opcdag{1} \\
\opcdag{2} \\
\vdots \\
\opcdag{\Nsites}
\end{array}
\right) 
\equiv \mathU_{\Gamma} \opbfPsi{}
\;,
\end{equation}
where $\opbfPsi{}$ is the Nambu vector of the original fermions, and the $2\Nsites\times 2\Nsites$ unitary matrix $\mathU_{\Gamma}$ has the following block form:
\begin{equation} \label{eqn:Ugamma_block}
\mathU_{\Gamma} = 
\left( 
\begin{array}{c|c}
\U_{\Gamma} & \V_{\Gamma}^* \\ \hline
\V_{\Gamma} & \U_{\Gamma}^* 
\end{array}
\right) \;.
\end{equation}
Here the $\Nsites\times \Nsites$ blocks $\U_{\Gamma}$ and $\V_{\Gamma}$ are given by:
\begin{equation} \label{eqn:Ugamma}
\U_{\Gamma} = \frac{i}{2} \left(
\begin{array}{cccccccc}
-1 & 0 & 0 & \cdots & 0 & 0 & +1 \\
+1 & 0 & 0 & \cdots & 0 & 0 & -1 \\
0 & -1 & 0 & \cdots & 0 & +1 & 0 \\
0 & +1 & 0 & \cdots & 0 & -1 & 0 \\
0 & 0 & \ddots & \vdots & \iddots & 0 & 0 \\
0 & 0 & \cdots & 0 & \cdots & 0 & 0
\end{array}
\right) \;, \hspace{5mm}
\V_{\Gamma} = \frac{i}{2} \left(
\begin{array}{cccccccc}
1 & 0 & 0 & \cdots & 0 & 0 & 1 \\
1 & 0 & 0 & \cdots & 0 & 0 & 1 \\
0 & 1 & 0 & \cdots & 0 & 1 & 0 \\
0 & 1 & 0 & \cdots & 0 & 1 & 0 \\
0 & 0 & \ddots & \vdots & \iddots & 0 & 0 \\
0 & 0 & \cdots & 2 & \cdots & 0 & 0
\end{array}
\right) \;.
\end{equation}
Using these new Dirac fermions we can write the target and driving Hamiltonians as follows:
\begin{equation}
\begin{aligned}
\Ho_z &= 2\sum_{j=1}^{(\Nsites-1)/2} J_j 
\Big(\Gammaopup_{2j} \Gammaopdown_{2j+1}
       - \Gammaopdown_{2j} \Gammaopup_{2j+1} \Big) 
+J_{\Nsites}\left(\Gammaopup_1\Gammaopdown_1
      - \Gammaopdown_1\Gammaopup_1\right) \;,
\\
\Ho_x &= -2h \sum_{j=1}^{(\Nsites-1)/2}
\Big( \Gammaopup_{2j-1} \Gammaopdown_{2j}
      - \Gammaopdown_{2j-1} \Gammaopup_{2j} \Big) 
-h \left( \Gammaopup_{\Nsites}\Gammaopdown_{\Nsites}      - \Gammaopdown_{\Nsites}\Gammaopup_{\Nsites}\right) \;,
\end{aligned}
\end{equation}
where $J_{\Nsites}=-J_f$ and $h=-J$.
Equivalently, in matrix form, with 
$\Gammaopbf^{\dagger} = \left(\Gammaopup_{1},\dots,\Gammaopup_{\Nsites},\Gammaopdown_{1}, \dots,\Gammaopdown_{\Nsites} \right)$:
\begin{eqnarray} \label{quadratic-H:eqn}
\Ho_{x/z} = \Gammaopbf^{\dagger} \, \mathH_{x/z} \, \Gammaopbf \;,
\end{eqnarray}
where $\mathH_{x/z}$ are $2\Nsites\times 2\Nsites$ Hermitian matrices of the form:
\begin{equation} \label{eqn:Hxz_Gamma}
\mathH_{x/z} = 
\left( 
\begin{array}{cc} \A_{x/z} & \mathbf{0} \\
\mathbf{0} & -\A_{x/z} \end{array} \right) \;.
\end{equation}
In turn, the $\Nsites\times \Nsites$ real symmetric matrices $\A_{x/z}$ can be written in terms of $2\times 2$ blocks:
\begin{equation} \label{eqn:AxAz_block}
\A_x = -h \left( 
\begin{array}{cc|cc|cc|cc|c} 
0 & 1 & 0 & 0 & 0 & 0 & \cdots & \cdots & 0\\
1 & 0 & 0 & 0 & 0 & 0 & \cdots & \cdots & 0\\
\hline
0 & 0 & 0 & 1 & 0 & 0 & \cdots & \cdots & 0\\
0 & 0 & 1 & 0 & 0 & 0 & \cdots & \cdots & 0\\
\hline
0 & 0 & 0 & 0 & \ddots & \ddots & 0 & 0 & 0\\
0 & 0 & 0 & 0 & \ddots & \ddots & 0 & 0 & 0\\
\hline
0 & 0 & \cdots & \cdots & 0 & 0 & 0 & 1 & 0\\
0 & 0 & \cdots & \cdots & 0 & 0 & 1 & 0 & 0\\
\hline
0 & 0 & 0 & 0 & \cdots & \cdots & 0 & 0 & 1
\end{array}
\right) \;, \hspace{5mm} 
\A_z = \left( 
\begin{array}{c|cc|cc|cc|cc} 
J_{\Nsites} & 0 & 0 & \cdots & \cdots & 0 & 0 & 0 & 0 \\
\hline
0 & 0 & J_1 & 0 & 0 &  \cdots & \cdots & 0 & 0\\
0 & J_1 & 0 & 0 & 0 &  \cdots & \cdots & 0 & 0\\
\hline
0 & 0 & 0 & 0 & J_2 & 0 & 0 & \cdots & \cdots \\
0 & 0 & 0 & J_2 & 0 & 0 & 0 & \cdots & \cdots \\
\hline
0 & 0 & 0 & 0 & 0 & \ddots & \ddots &  0 & 0\\
0 & 0 & 0 & 0 & 0 & \ddots & \ddots &  0 & 0\\
\hline
0 & 0 & 0 & \cdots & \cdots & 0 & 0 & 0 & J_w \\
0 & 0 & 0 & \cdots & \cdots & 0 & 0 & J_w & 0 \\
\end{array}
\right) \;.
\end{equation}

To get the final variational energy, we start from the discrete-time expression:
\begin{align} \label{eqn:EP_2}
E_{\Ptrot}(\btheta) 
&\equiv \braket{\psi_{\Ptrot}(\btheta) | \Ho_z | \psi_{\Ptrot}(\btheta)} 
= \braket{\psi_0| \Uo_{\text{ev}}^{\dagger}(\btheta) \, \Ho_z \, \Uo_{\text{ev}}(\btheta) | \psi_{0}} \;,
\end{align}
where 
\begin{equation}
\Uo_{\text{ev}}(\btheta) =  \Uo_{\Ptrot}(\theta^x_\Ptrot,\theta^z_\Ptrot) \, 
\cdots \, \Uo_1(\theta^x_1,\theta^z_1) \;,
\end{equation}
is the digitized-and-Trotterized evolution operator. 
Next, we move to the Heisenberg picture for the operators. 
Considering that, see Eq.~\eqref{eqn:Hxz_Gamma},  $\Ho_z=\Gammaopbf^{\dagger}\, \mathH_z \, \Gammaopbf$, then 
\begin{align} \label{eqn:UdagHzU}
\Uo_{\text{ev}}^{\dagger}(\btheta) \, \Ho_z \, \Uo_{\text{ev}}(\btheta)
&= \sum_{j,j'=1}^{2\Nsites} \Gammaopbf^{\dagger}_{j'\Heis}(\btheta) \, \big(\mathH_z\big)_{j'j} \, \Gammaopbf_{j\Heis}(\btheta)  %
= \Gammaopbf^{\dagger}_{\Heis}(\btheta) \, \mathH_z \, \Gammaopbf_{\Heis}(\btheta) \;, 
\end{align}
where 
\begin{equation}
\Gammaopbf_{j\Heis}(\btheta) = 
\Uo_{\text{ev}}^{\dagger}(\btheta) \, \Gammaopbf_{j} \, \Uo_{\text{ev}}(\btheta) \;,
\end{equation}
is the Heisenberg form of the (Nambu) symmetry-adapted fermionic operators. 
In precise analogy with the time-dependent Bogoljubov-de Gennes (BdG) theory~\cite{mbeng2024quantum} --- viewing the digital evolution as a particular stepwise time-dependence of the Hamiltonian --- we deduce that
\begin{equation} \label{eqn:Gamma_Heis}
\Gammaopbf_{\Heis}(\btheta) = \mathU_{\text{ev}}(\btheta) \, \Gammaopbf \;,
\end{equation}
where $\mathU_{\text{ev}}(\btheta)$ is a 
$2\Nsites\times 2\Nsites$ unitary operator solving the time-dependent BdG equations.
The result can be cast in the form:
\begin{equation} \label{eq:evolutionboldtheta}
\mathU_{\text{ev}}(\btheta) =
\mathU_\Ptrot(\theta^x_\Ptrot,\theta^z_\Ptrot) \, \dots \, 
\mathU_1(\theta^x_1,\theta^z_1) \,
\end{equation}
where, for $p=1,\cdots,\Ptrot$:
\begin{equation} \label{eqn:U_nambu_app}
\mathU_p(\theta^x_p,\theta^z_p) 
= \nep^{-2i \theta^x_p \, \mathH_x} 
\nep^{-2i \theta^z_p \, \mathH_z} \;.
\end{equation}
Notice that, due to the block form of $\mathH_{x/z}$ [see Eqs.~\eqref{eqn:Hxz_Gamma} and~\eqref{eqn:AxAz_block}], the various $\mathU_p$ can be calculated analytically.
In particular, the exponentials of the blocks $\A_x$ and $\A_z$ are given by
\begin{equation} \label{eqn:expAx}
\nep^{-2i\theta^x_p \A_x} = \left( 
\begin{array}{cc|cc|cc|c} 
\cos(2h\theta^x_p) & i\sin(2h\theta^x_p) & 0 & 0 & \cdots & \cdots & 0\\
i\sin(2h\theta^x_p) & \cos(2h\theta^x_p) & 0 & 0 & \cdots & \cdots & 0\\
\hline
0 & 0 & \ddots & \ddots & 0 & 0 & 0\\
0 & 0 & \ddots & \ddots & 0 & 0 & 0\\
\hline
0 & 0 & \cdots & \cdots &  \cos(2h\theta^x_p) & i\sin(2h\theta^x_p) & 0\\
0 & 0 & \cdots & \cdots &  i\sin(2h\theta^x_p) & \cos(2h\theta^x_p) & 0\\
\hline
0 & 0  & \cdots & \cdots & 0 & 0 & \cos(2h\theta^x_p) + i\sin(2h\theta^x_p) 
\end{array}
\right) \;, 
\end{equation}
and
\begin{equation}
\nep^{-2i\theta^z_p \A_z} = \left( 
\begin{array}{c|cc|cc|cc} 
\cos(2J_{\Nsites}\theta^z_p) - i\sin(2J_{\Nsites}\theta^z_p) & 0 & 0 & 0 & 0 & 0 & 0 \\
\hline
0 & \cos(2J\theta^z_p) & -i\sin(2J\theta^z_p) &  \cdots & \cdots & 0 & 0\\
0 & -i\sin(2J\theta^z_p) & \cos(2J\theta^z_p) &  \cdots & \cdots & 0 & 0\\
\hline
0 & 0 & 0 & \ddots & \ddots &  0 & 0\\
0 & 0 & 0  & \ddots & \ddots &  0 & 0\\
\hline
0 & 0 & 0  & 0 & 0 & \cos(2J_w\theta^z_p) & -i\sin(2J_w\theta^z_p) \\
0 & 0 & 0  & 0 & 0 & -i\sin(2J_w\theta^z_p) & \cos(2J_w\theta^z_p) \\
\end{array}
\right) \;.
\end{equation}
These expressions make it particularly easy to calculate the various ingredients appearing in the time-evolution of the state, including their derivatives with respect to the parameters $\btheta$.  
The total $\mathU_{\text{ev}}(\btheta)$ in Eq.~\eqref{eq:evolutionboldtheta} is also in block form:
\begin{equation} \label{eqn:Uev_block}
\mathU_{\text{ev}}(\btheta) = 
\left( 
\begin{array}{cc} \U_{\text{ev}}(\btheta) & \mathbf{0} \\
\mathbf{0} & \U_{\text{ev}}^*(\btheta) 
\end{array} \right) \;.
\end{equation}
The block forms of $\mathU_{\text{ev}}$ and $\mathH_z$ in Eqs.~\eqref{eqn:Hxz_Gamma} and ~\eqref{eqn:Uev_block} imply that:
\begin{equation} \label{eqn:UdagHzUmath}
\mathU_{\text{ev}}^{\dagger}(\btheta) \, \mathH_z \, \mathU_{\text{ev}}(\btheta) = 
\left( 
\begin{array}{cc} \A_{z}(\btheta) & \mathbf{0} \\
\mathbf{0} & -\A_{z}^*(\btheta) 
\end{array} \right) 
\end{equation}
where $\A_z(\btheta)$ denotes the Hermitian matrix obtained by transforming $\A_z$ with the unitary evolution block $\U_{\text{ev}}(\btheta)$: 
\begin{equation}
\A_z(\btheta) = \U_{\text{ev}}^{\dagger}(\btheta) \, \A_z \, \U_{\text{ev}}(\btheta) \;.
\end{equation}

The final energy in Eq.~\eqref{eqn:EP_2} can therefore be expressed, using Eqs.~\eqref{eqn:UdagHzU} and ~\eqref{eqn:Gamma_Heis}, as:
\begin{align} \label{eqn:EP_3}
E_{\Ptrot}(\btheta) = \langle \psi_0 |
\Gammaopbf^{\dagger}_{\Heis}(\btheta) \, \mathH_z \, \Gammaopbf_{\Heis}(\btheta) |\psi_0\rangle %
&= \langle \psi_0 |
\Gammaopbf^{\dagger}  \, \mathU_{\text{ev}}^{\dagger}(\btheta) \, \mathH_z \, \mathU_{\text{ev}}(\btheta) \, \Gammaopbf |\psi_0\rangle \nonumber \\
&= \sum_{j,j'=1}^{2\Nsites} 
\big( \mathU_{\text{ev}}^{\dagger}(\btheta) \, \mathH_z \, \mathU_{\text{ev}}(\btheta) \big)_{j,j'}
\langle \psi_0 | \Gammaopbf^{\dagger}_j \Gammaopbf_{j'}|\psi_0\rangle 
\;.
\end{align}
To continue, we use Eq.~\eqref{eqn:Gamma_vs_Psi} to calculate the matrix
\begin{align}
\mathGamma_{j,j'} \equiv \langle \psi_0 | \Gammaopbf^{\dagger}_j \Gammaopbf_{j'}|\psi_0\rangle 
&= \sum_{l,l'=1}^{2\Nsites} \big(\mathU^{\dagger}_{\Gamma}\big)_{l,j} \, 
\big(\mathU_{\Gamma}\big)_{j',l'} \,
\langle \psi_0 | \opbfPsidag{l} \opbfPsi{l'}|\psi_0\rangle 
= \sum_{l=1}^{\Nsites} 
\big(\mathU_{\Gamma}\big)_{j',l} \, \big(\mathU^{\dagger}_{\Gamma}\big)_{l,j} \;,
\end{align}
where we used $\langle \psi_0 | \opbfPsidag{l} \opbfPsi{l'}|\psi_0\rangle = \langle \psi_0 | \opcdag{l} \opc{l'}|\psi_0\rangle = \delta_{l,l'} $ for $l,l' \le \Nsites$,
as appropriate for the fully occupied initial state $|\psi_0\rangle = \prod_{j=1}^{\Nsites} \opcdag{j}|0\rangle$ 
relevant for $h<0$.
In turn, the block form of $\mathU_{\Gamma}$ in Eq.~\eqref{eqn:Ugamma_block} implies that for $j,j'\in \{1,\cdots,\Nsites\}$:
\begin{align}
\mathGamma_{j,j'} = \big( \U_{\Gamma} \U_{\Gamma}^{\dagger} \big)_{j',j} \;, \hspace{10mm}
\mathGamma_{\Nsites+j,\Nsites+j'} = \big( \V_{\Gamma} \V_{\Gamma}^{\dagger} \big)_{j',j} \;.
\end{align}
By inserting these expressions into Eq.~\eqref{eqn:EP_3} we deduce that:
\begin{align} \label{eqn:EP_4}
E_{\Ptrot}(\btheta) = \sum_{j,j'=1}^{2\Nsites} 
\big( \mathU_{\text{ev}}^{\dagger}(\btheta) \, \mathH_z \, \mathU_{\text{ev}}(\btheta) \big)_{j,j'}
\langle \psi_0 | \Gammaopbf^{\dagger}_j \Gammaopbf_{j'}|\psi_0\rangle 
&= \sum_{j,j'=1}^{\Nsites} \Big(
\big( \A_z(\btheta) \big)_{j,j'} \mathGamma_{j,j'} -
\big( \A_z^*(\btheta) \big)_{j,j'} \mathGamma_{\Nsites+j,\Nsites+j'} 
\Big) \nonumber \\
&= \Tr \Big( \A_z(\btheta) \U_{\Gamma} \U_{\Gamma}^{\dagger} 
- \A_z^*(\btheta) \V_{\Gamma} \V_{\Gamma}^{\dagger} \Big)
 \;.
\end{align}
The explicit expressions for $\U_{\Gamma}$ and $\V_{\Gamma}$ in Eq.~\eqref{eqn:Ugamma} imply that:
\begin{equation}
\U_{\Gamma} \U_{\Gamma}^{\dagger} = \frac{1}{2}\left( 
\begin{array}{rr|rr|cc|rr|c} 
1 & -1 & 0 & 0 & 0 & 0 & \cdots & \cdots & 0\\
-1 & 1 & 0 & 0 & 0 & 0 & \cdots & \cdots & 0\\
\hline
0 & 0 & 1 & -1 & 0 & 0 & \cdots & \cdots & 0\\
0 & 0 & -1 & 1 & 0 & 0 & \cdots & \cdots & 0\\
\hline
0 & 0 & 0 & 0 & \ddots & \ddots & 0 & 0 & 0\\
0 & 0 & 0 & 0 & \ddots & \ddots & 0 & 0 & 0\\
\hline
0 & 0 & \cdots & \cdots & 0 & 0 & 1 & -1 & 0\\
0 & 0 & \cdots & \cdots & 0 & 0 & -1 & 1 & 0\\
\hline
0 & 0 & 0 & 0 & \cdots & \cdots & 0 & 0 & 0
\end{array}
\right) \;, \hspace{2mm}
\V_{\Gamma} \V_{\Gamma}^{\dagger} = \frac{1}{2}
\left( 
\begin{array}{rr|rr|cc|rr|c} 
1 & 1 & 0 & 0 & 0 & 0 & \cdots & \cdots & 0\\
1 & 1 & 0 & 0 & 0 & 0 & \cdots & \cdots & 0\\
\hline
0 & 0 & 1 & 1 & 0 & 0 & \cdots & \cdots & 0\\
0 & 0 & 1 & 1 & 0 & 0 & \cdots & \cdots & 0\\
\hline
0 & 0 & 0 & 0 & \ddots & \ddots & 0 & 0 & 0\\
0 & 0 & 0 & 0 & \ddots & \ddots & 0 & 0 & 0\\
\hline
0 & 0 & \cdots & \cdots & 0 & 0 & 1 & 1 & 0\\
0 & 0 & \cdots & \cdots & 0 & 0 & 1 & 1 & 0\\
\hline
0 & 0 & 0 & 0 & \cdots & \cdots & 0 & 0 & 2
\end{array}
\right) \;.
\end{equation}
The fact that $\V_{\Gamma} \V_{\Gamma}^{\dagger}$ is symmetric, $\A_z(\btheta)$ is Hermitian, and the invariance of the trace under transposition and cyclic permutations imply that:
\begin{align}
  \Tr \Big( \A_z^*(\btheta) \, \V_{\Gamma} \V_{\Gamma}^{\dagger} \Big)  &= \Tr \Big(  \big(\V_{\Gamma} \V_{\Gamma}^{\dagger} \big)^{\transpose} \A_z^{\dagger}(\btheta) \Big) 
  = \Tr \Big( \A_z(\btheta) \, \V_{\Gamma} \V_{\Gamma}^{\dagger} \Big) \;.
\end{align}
Hence, we conclude that
\begin{align} \label{eqn:EP_5}
E_{\Ptrot}(\btheta) = \Tr \Big( \A_z(\btheta) \, \U_{\Gamma} \U_{\Gamma}^{\dagger} 
- \A_z^*(\btheta) \, \V_{\Gamma} \V_{\Gamma}^{\dagger} \Big)
&= \Tr \Big( \A_z(\btheta) \,  
\big( \U_{\Gamma} \U_{\Gamma}^{\dagger} 
- \V_{\Gamma} \V_{\Gamma}^{\dagger}\big) \Big) %
= \Tr \Big( \U_{\text{ev}}^{\dagger}(\btheta) \, \A_z \,  \U_{\text{ev}}(\btheta)\, \G \Big) \;,
\end{align}
where the symmetric matrix $\G$ is given by:
\begin{align}
\G = \U_{\Gamma} \U_{\Gamma}^{\dagger} 
- \V_{\Gamma} \V_{\Gamma}^{\dagger} 
= -
\left( 
\begin{array}{rr|rr|cc|rr|c} 
0 & 1 & 0 & 0 & 0 & 0 & \cdots & \cdots & 0\\
1 & 0 & 0 & 0 & 0 & 0 & \cdots & \cdots & 0\\
\hline
0 & 0 & 0 & 1 & 0 & 0 & \cdots & \cdots & 0\\
0 & 0 & 1 & 0 & 0 & 0 & \cdots & \cdots & 0\\
\hline
0 & 0 & 0 & 0 & \ddots & \ddots & 0 & 0 & 0\\
0 & 0 & 0 & 0 & \ddots & \ddots & 0 & 0 & 0\\
\hline
0 & 0 & 0 & 0 & \cdots & \cdots & 0 & 1 & 0\\
0 & 0 & 0 & 0 & \cdots & \cdots & 1 & 0 & 0\\
\hline
0 & 0 & 0 & 0 & \cdots & \cdots & 0 & 0 & 1
\end{array}
\right) \;.
\end{align}

\section{The Trotter error} \label{app:trotter}
To assess the effect of the digital approximation on the continuous-time dynamics, we compared the residual energy obtained from the exact continuous-time evolution, $\epsilon_{\text{ct}}(\tau)$, with the corresponding digitized residual energy $\epsilon_{\Ptrot}(\tau)$ (see Eq.~\eqref{eqn:res_energy}) as a function of the number of Trotter steps $\Ptrot$. To do this, we fixed $\tau$ and then optimized the protocols for different values of $\Ptrot$, obtaining different values for $\epsilon_{\Ptrot}(\tau)$. For each of these $\Ptrot$, we used the variational parameters $\C^{\opt}$ to obtain $s^{\opt}(t)$ and $\epsilon_{\text{ct}}(\tau)$, and then compared the two different residual energies. A representative example is shown in Fig.~\ref{fig:residualenergyscaling} for $\tau=40$, $\Nsites=13$, and $J_f=0.45$.

As expected, the digitized result converges to the continuous-time value as $\Ptrot$ increases. The difference $\Delta \epsilon = |\epsilon_{\mathrm{ct}}-\epsilon_{\Ptrot}|$ decreases approximately as a power law compatible with $\Ptrot^{-2}$ in the explored range, consistently with the use of a first-order Trotter splitting for the unitary evolution over a fixed total time.

This analysis justifies our working choice $\Ptrot=10  *  \mathrm{int}(\tau)$, corresponding to $\deltat=0.1$, which provides a good compromise between numerical accuracy and computational cost throughout the paper.

\begin{figure*}[htp]
\centering
\includegraphics[width=80mm]{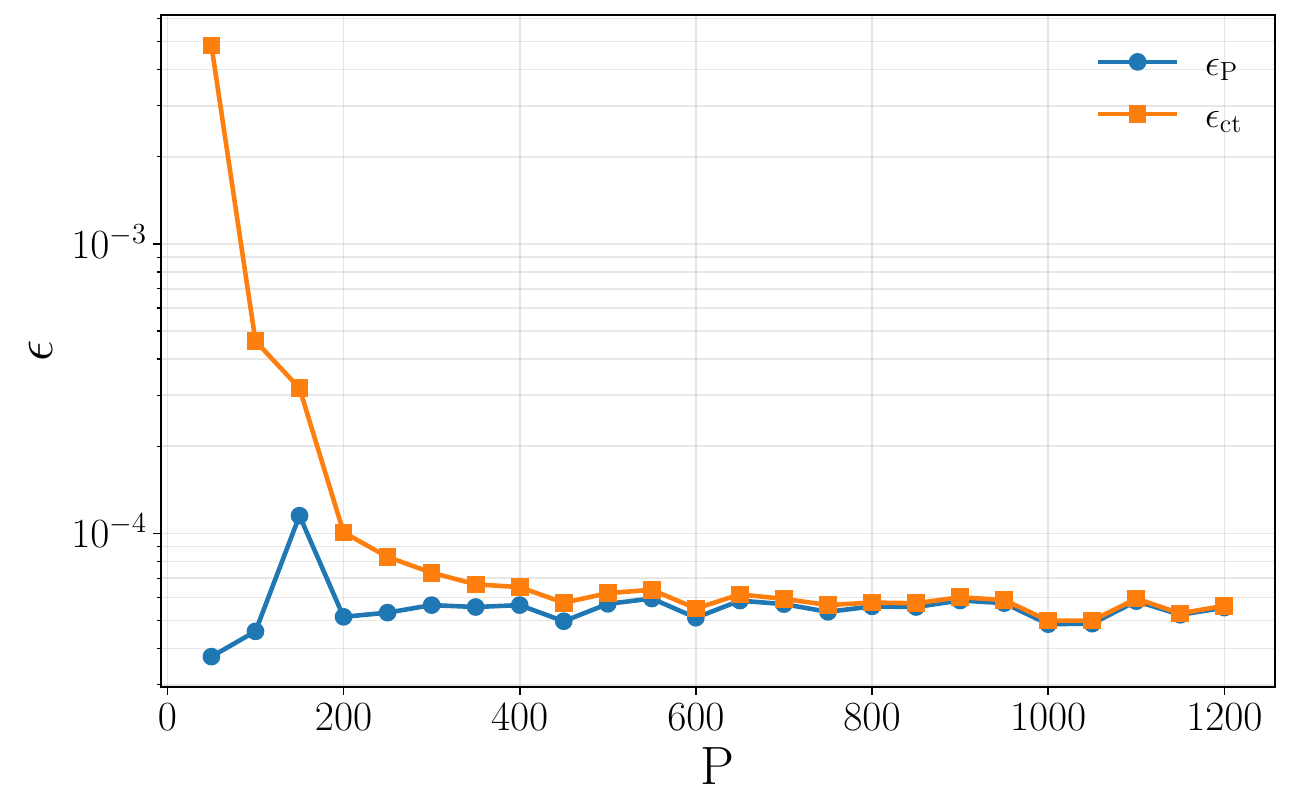}
\includegraphics[width=80mm]{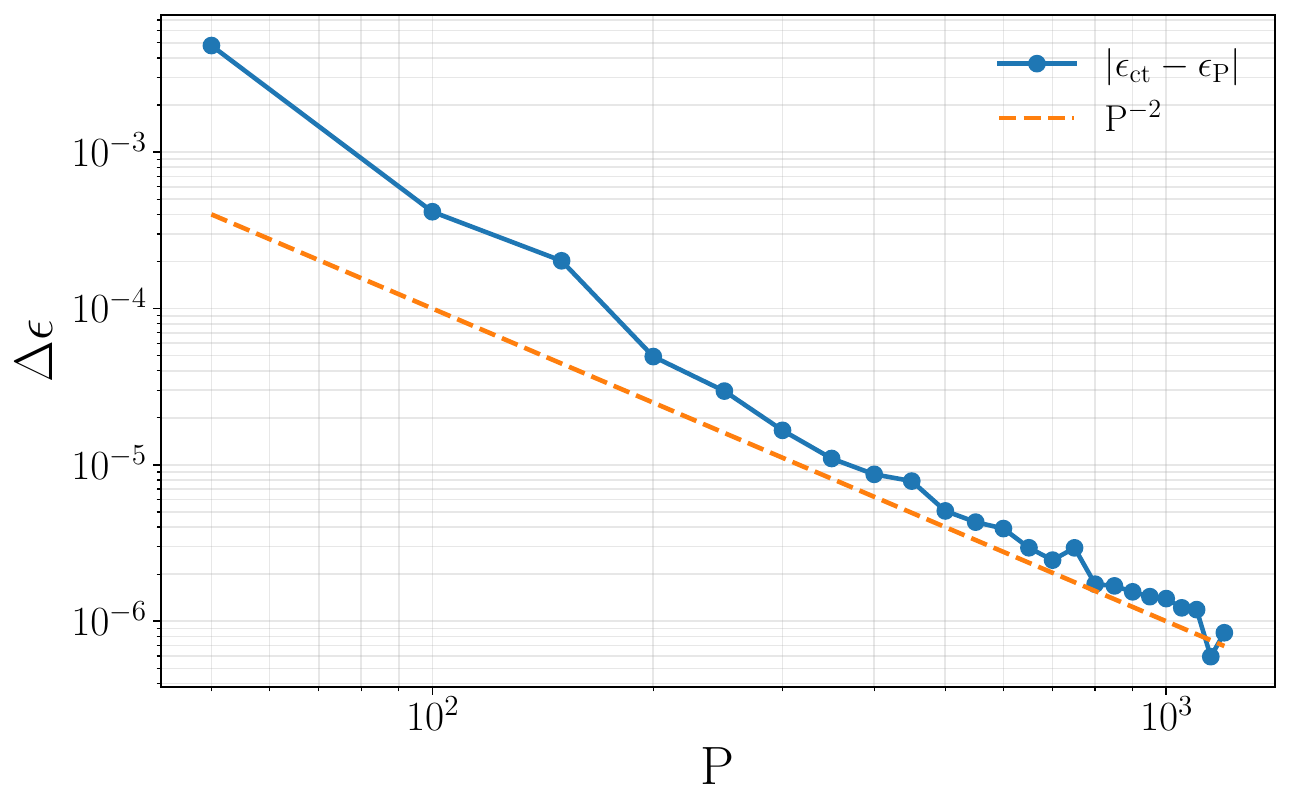}
\caption{Comparison between continuous-time residual energy $\epsilon_{\text{ct}}(\tau)$ and the digitized residual energy $\epsilon_{\Ptrot}(\tau)$ versus the number of steps $\Ptrot$ in which the dynamics is Trotterized. The plots correspond to $\tau = 40$, $\Nsites = 13$, $J_f = 0.45J$. 
For $\Ptrot = 400$, $\Delta\epsilon \equiv\left|\epsilon_{\text{ct}} - \epsilon_{\Ptrot}\right| \approx 9\times10^{-6} < \epsilon_\Ptrot \approx 5.7 \times 10^{-5}.$
}
\label{fig:residualenergyscaling}
\end{figure*}

\section{Adiabatic gauge potential} \label{app:counter-diabatic}
The lowest-order approximation for the adiabatic gauge potential reads:
\begin{equation}
\hat{\calA}^{\LO}(s) = i \alpha_1(s) [\Ho(s),\partial_s \Ho(s)] \;.
\end{equation}
where the function $\alpha_1(s)$ is found by minimizing 
an {\em action}~\cite{KOLODRUBETZ20171}:
\begin{align}
    S(\alpha_1) &= \Tr \Big( \partial_s \Ho(s) + \frac{\alpha_1}{\hbar} \big[ \Ho(s), [\Ho(s), \partial_s \Ho(s) \big]\big] \Big)^2 \nonumber \\
    &= \Tr \big(\partial_s \Ho(s)\big)^2 
    + \frac{\alpha_1^2}{\hbar^2} \Tr \big(\big[ \Ho(s), \big[\Ho(s), \partial_s \Ho(s) \big]\big]\big)^2 
    + 2 \frac{\alpha_1}{\hbar} \Tr \big( (\partial_s \Ho(s)) \big[ \Ho(s), \big[\Ho(s), \partial_s \Ho(s) \big]\big] \big) \;.
    \nonumber
\end{align}
The action is therefore a quadratic function of $\alpha_1$.
Setting to 0 the derivative of $S(\alpha_1)$ we get an expression for the optimal $\alpha_1$. 
Using simple properties of the trace we get:
\begin{equation} \label{eqn:alpha_solution}
    \alpha_1^{\opt}(s) 
    = 
    - \frac{\hbar \, \Tr (i[\Ho(s),\partial_s\Ho])^2}{\Tr([\Ho(s),[\Ho(s),\partial_s\Ho]])^2} 
    \defuguale - \alpha(s) \;,
\end{equation}
where the last expression follows from observing that the square of an anti-Hermitian operator is negative: the quantity $\alpha(s)$ is, by definition, \textbf{positive}. To the lowest order, the counter-diabatic Hamiltonian reads, therefore:
\begin{equation}
    \Ho_{\CD}^{\LO}(s,\dot{s}) = \Ho(s) -i \dot{s} \alpha(s) \, [\Ho(s),\partial_s \Ho(s)] \;.
\end{equation}
Recalling that $\Ho(s)=s\Ho_z+(1-s)\Ho_x$, the relevant ingredients are:
\begin{align}
[\Ho(s),\partial_s \Ho(s)] &= [\Ho_x,\Ho_z] \nonumber \\
[\Ho(s),[\Ho(s),\partial_s \Ho(s)]] &= s [\Ho_z,[\Ho_x,\Ho_z]] + (1-s) [\Ho_x,[\Ho_x,\Ho_z]]
\end{align}
To calculate the various commutators in a way that makes the Jordan-Wigner transformation easier, we prefer to make a spin rotation by $\pi/2$ around the $y$ axis such that $\PauliSigma^x\to \PauliSigma^z$ and $\PauliSigma^z\to -\PauliSigma^x$, so that:
\begin{align}
    \Ho_z \to -\sum_{j=1}^{\Nsites} J_j \PauliSigma^x_j \PauliSigma^x_{j+1} \;, \hspace{10mm}
    \Ho_x \to -h \sum_{j=1}^{\Nsites} \PauliSigma^z_j \;.
\end{align}
Hence, using the angular momentum algebra, we deduce that:
\begin{align}
i [\Ho_x , \Ho_z] 
&= -2h \sum_{j=1}^{\Nsites}  J_j \Big( \PauliSigma^x_j \PauliSigma^y_{j+1} + \PauliSigma^y_j \PauliSigma^x_{j+1} \Big) \;, \nonumber \\
[\Ho_x,[\Ho_x,\Ho_z]] 
&= 8 h^2 \sum_{j=1}^{\Nsites}  J_j \Big( \PauliSigma^y_j \PauliSigma^y_{j+1} - \PauliSigma^x_j \PauliSigma^x_{j+1} \Big) \;, \nonumber \\
[\Ho_z,[\Ho_x,\Ho_z]] 
&= 8h \sum_{j=1}^{\Nsites}  J_{j} J_{j+1} \, \PauliSigma^x_{j} \PauliSigma^z_{j+1} \PauliSigma^x_{j+2} 
+4h \sum_{j=1}^{\Nsites} \Big( J_{j-1}^2 + J_j^2 \Big) \PauliSigma^z_j \;.
\end{align}

The function $\alpha(s)$ 
\begin{equation}
\alpha(s) = \frac{ \hbar\, \Tr \big(i[\Ho_x,\Ho_z]\big)^2}{\Tr\big( (1-s) [\Ho_x,[\Ho_x,\Ho_z]] + s [\Ho_z,[\Ho_x,\Ho_z]] \big)^2} \;.
\end{equation}
involves traces of Pauli operators, and can be calculated explicitly. 
The crucial property is that unpaired Pauli matrices do not contribute, and that the square of a Pauli matrix gives the identity, hence a factor $2$ under trace. 
The numerator 
reads:
\[
\Tr \big(i[\Ho_x,\Ho_z]\big)^2 = 2^{\Nsites} 8 h^2 \sum_j J_j^2 \;.
\]
The denominator reads:
\begin{equation}
2^{\Nsites} \Big( 128 \, h^4 (1-s)^2 \sum_j J_j^2 + s^2 \big( 96 h^2 \sum_j J_j^2 J_{j+1}^2 + 32 h^2 \sum_j J_j^4 \big)\Big) \;.
\end{equation}
Putting these ingredients together we get:
\begin{equation}
    \alpha(s) = \frac{\sum_j J_j^2}{16 (1-s)^2 \sum_j J_j^2 + s^2( 12 \sum_j J_j^2 J_{j+1}^2 + 4 \sum_j J_j^4 ) } \;.
\end{equation}
With the choice of couplings for the frustrated ring model we have:
\[
\left\{
\begin{array}{rcl} 
\sum_j J_j^2 &=& (\Nsites-3) J^2 + 2 J_w^2 + J_f^2 \vspace{3mm} \\
\sum_j J_j^4 &=& (\Nsites-3) J^4 + 2 J_w^4 + J_f^4 \vspace{3mm} \\
\sum_j J_j^2 J_{j+1}^2 &=& (\Nsites-5) J^4 + 2 J^2 J_w^2 + 2 J^2 J_f^2 
\end{array} \;.
\right. 
\]

Let us now use Jordan-Wigner to rewrite these terms with spinless fermions. 
We have:
\begin{equation}
\hat{\calA}^{\LO}(s) = - i \alpha(s)[\Ho_x , \Ho_z] 
= - 2 \alpha(s)\sum_{j=1}^{\Nsites} J_j \Big( \PauliSigma^x_j \PauliSigma^y_{j+1} + \PauliSigma^y_j \PauliSigma^x_{j+1} \Big)
= - 4i \alpha(s) \sum_{j=1}^{\Nsites} J_j \Big( \opcdag{j} \opcdag{j+1} - \opc{j+1} \opc{j} \Big) \;.
\end{equation}

Now we need to apply the transformations in Eqs.~\eqref{eq:gammaoddj}-\eqref{eq:gammaevenj}. 
Defining $\hat{\calA}^{\LO}(s) = \alpha(s) \hat{H}_{xz}$, and $\ell=(\Nsites-1)/2$, so that $\Nsites=2\ell +1$, we get the following expression for $\hat{H}_{xz}=-i[\Ho_x , \Ho_z]$:
\begin{align}
\hat{H}_{xz}   &= 
4 i \sum_{j=1}^{\ell} J_j 
\left( \hat{\Gamma}^{+}_{2j+1}\hat{\Gamma}^{-}_{2j-1} - \hat{\Gamma}^{+}_{2j-1}\hat{\Gamma}^{-}_{2j+1} \right) 
+ 4i \sum_{j=1}^{\ell-1} J_j 
\left( \hat{\Gamma}^{+}_{2j}\hat{\Gamma}^{-}_{2j+2} - \hat{\Gamma}^{+}_{2j+2}\hat{\Gamma}^{-}_{2j} \right) 
\quad \nonumber \\
&+ 4i J_{\ell} 
\left( \hat{\Gamma}^{+}_{2\ell}\hat{\Gamma}^{-}_{2\ell+1} - \hat{\Gamma}^{+}_{2\ell+1}\hat{\Gamma}^{-}_{2\ell} \right) 
+ 4i  J_{\Nsites} 
\left( \hat{\Gamma}^{+}_{1}\hat{\Gamma}^{-}_{2} - \hat{\Gamma}^{+}_{2}\hat{\Gamma}^{-}_{1} \right) \,,
\end{align}
where, we should recall that $J_{\ell}=J_w$ and $J_{\Nsites}=-J_f$.
We can rewrite $\hat{H}_{xz}$ in the Nambu \cite{mbeng2024quantum} formalism as
\begin{equation}
    \hat{H}_{xz} = \Gammaopbf^{\dagger}{} \, \mathH_{xz} \, \Gammaopbf{} \;,
\end{equation}
where $\mathH_{xz}$,
\begin{equation}
\mathH_{xz} = 
\left( \begin{array}{cc} \A_{xz} & \mathbf{0} \\
        \mathbf{0} & -\A_{xz} \end{array} \right) \;.    
\end{equation}
The Hermitian and purely imaginary matrix $\A_{xz}$ is defined as follows:
\begin{align}
\left(\A_{xz}\right)_{1,2} &= -2i\,J_f, &
\left(\A_{xz}\right)_{2,1} &= 2i\,J_f, \\
\left(\A_{xz}\right)_{\Nsites-1,\Nsites} &= 2i\,J_{w}, &
\left(\A_{xz}\right)_{\Nsites,\Nsites-1} &= -2i\,J_{w}.
\end{align}
Moreover, the third diagonals (for $j=1,\dots,\ell$) are given by
\begin{align}
\left(\A_{xz}\right)_{2j-1,2j+1} &= -\,2i\,J_{j}, & \text{if}\;\; 2j+1 \leq \Nsites, \nonumber \\
\left(\A_{xz}\right)_{2j,2j+2}  &= -\,2i\,J_{j}, & \text{if} \;\; 2j+2 \leq \Nsites \;.
\end{align}
Hence, even by adding the counter-diabatic gauge potential, we can restrict the dynamics to the $\Nsites \times \Nsites$ matrix $\A_{xz}$.

\end{widetext}

\bibliography{biblio}

\end{document}